\documentclass{IEEEoj}
\usepackage{cite}
\usepackage{amsmath,amssymb,amsfonts}
\usepackage{algorithmic}
\usepackage{graphicx,color}
\usepackage{textcomp}
\usepackage{url}
\usepackage{soul}
\usepackage{tcolorbox}
\def\BibTeX{{\rm B\kern-.05em{\sc i\kern-.025em b}\kern-.08em
    T\kern-.1667em\lower.7ex\hbox{E}\kern-.125emX}}
\AtBeginDocument{\definecolor{ojcolor}{cmyk}{0.93,0.59,0.15,0.02}}
\def\OJlogo{\vspace{-4pt}\hskip-4pt\includegraphics[height=18pt]{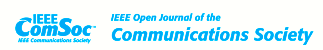}}

\newcommand{\system}{C-RE-ACT}

\newcommand{\fnode}{\textit{F}-Node}
\newcommand{\uesnode}{\textit{UEs}-Node}

\usepackage{graphicx}
\usepackage{subcaption}
\usepackage{float}
\usepackage{booktabs}
\let\labelindent\relax          
    \usepackage{enumitem}
    \newlist{promptitems}{itemize}{3}
    \setlist[promptitems]{label=\textendash,leftmargin=1.1em,
                          itemsep=1pt,topsep=2pt,parsep=0pt}
    \newlist{promptsteps}{enumerate}{3}
    \setlist[promptsteps]{label=\arabic*.,leftmargin=1.4em,
                          itemsep=3pt,topsep=3pt,parsep=0pt}
    \newcommand{\code}[1]{\texttt{#1}}
    \newcommand{\phead}[1]{\par\smallskip\noindent\textbf{#1}\par\nobreak\smallskip}
\usepackage{soul}
\usepackage{tcolorbox}

\newtcolorbox[auto counter]{mydefinition}[2][]{
colback=blue!5!white,
colframe=blue!75!black, 
fonttitle=\bfseries, 
title=Definition \thetcbcounter: {#2}, #1
}

\newtcolorbox[auto counter]{myllmoutputbox}[2][]{
    title={LLM Output \thetcbcounter: {#2}},
    label={#1},
    colback=yellow!10,
    fonttitle=\bfseries\fontsize{8}{10}\selectfont,
    fontupper=\fontfamily{ptm}\fontsize{8}{10}\selectfont
}

\newtcolorbox[auto counter]{myllminputbox}[2][]{
    title={LLM Prompt \thetcbcounter: {#2}},
    label={#1},
    colback=green!10,
    fonttitle=\bfseries\fontsize{8}{10}\selectfont,
    fontupper=\fontfamily{ptm}\fontsize{8}{10}\selectfont
}

\newcommand{\rev}[1]{\textcolor{black}{#1}}

\usepackage{etoolbox}

\makeatletter
\renewcommand{\p@subsection}{\thesection-}
\renewcommand{\p@subsubsection}{\thesection-\thesubsection}
\pretocmd{\@maketitle}{%
  \vspace{-5mm}
  \begingroup
  \setlength{\fboxsep}{2pt}%
  \setlength{\fboxrule}{0.4pt}%
  \noindent\fbox{%
    \begin{minipage}{\dimexpr\textwidth-2\fboxsep-2\fboxrule\relax}
      \footnotesize
      \parindent=0pt \leftskip=0pt \rightskip=0pt
      \parfillskip=0pt plus 1fil
      \centering
      This work has been submitted to the IEEE for possible publication.
      Copyright may be transferred without notice, after which this version
      may no longer be accessible.\par
    \end{minipage}}%
  \vspace{-3mm}
  \endgroup
}{}{\typeout{NOTICE PATCH FAILED}}
\let\@revised\@empty
\makeatother

\begin{document}
\bstctlcite{BSTcontrol}


\title{C-RE-ACT: Causal RE-ACTing Agent for O-RAN Forensic Triage}

\author{ 
    J. Xavier Salvat Lozano \IEEEauthorrefmark{2,3}, 
    Gines Garcia-Aviles \IEEEauthorrefmark{1},
    Pau Baguer\IEEEauthorrefmark{1},
    and Xavier Costa-Pérez \IEEEauthorrefmark{1,2,4} \IEEEmembership{(Senior Member, IEEE)}
}

\affil{i2CAT Foundation, 08034 Barcelona, Spain}
\affil{NEC Laboratories Europe GmbH, 69115 Heidelberg, Germany}
\affil{Universitat Autonoma de Barcelona, 08193 Barcelona, Spain}
\affil{ICREA, 08010 Barcelona, Spain}

\corresp{CORRESPONDING AUTHOR: Gines Garcia-Aviles (e-mail: gines.garcia@i2cat.net).}

\authornote{This work was supported in part by the ORIGAMI Project under Grant 101139270; in part by the CERCA Programme from the Generalitat de Catalunya through the ICREA programme; and in part co-funded by Departament de Recerca i Universitats under the frame of the Resilicat project.}

\markboth{Preparation of Papers for IEEE OPEN JOURNALS}{Author \textit{et al.}}

\begin{abstract}

The shift to Open RAN (O-RAN) architectures marks a turning point in cellular security: disaggregated components and open interfaces \rev{broaden the attack surface and introduce new} vectors for misconfiguration, failure, and exploitation. \rev{Performance-degradation attacks constitute the largest class of threats cataloged by O-RAN Alliance Working Group 11.} \rev{They} induce packet losses and latency spikes that are hard to distinguish from operational events such as misconfigurations, transient congestion, or software regressions. \rev{Upon detecting an incident}, support engineers must \rev{therefore} rapidly \rev{decide} whether to route the ticket to network maintenance or escalate it to security operations\rev{, a critical human-in-the-loop bottleneck in incident response}. We introduce \system{} (Causal RE-ACTing agent), an automated agentic triage framework \rev{that generates incident reports localizing the degradation and recommending diagnostic actions}. \system{} \rev{first constructs} a Weighted Directed Acyclic Graph (WDAG) over O-RAN metrics using the Structural Agnostic Model (SAM). \rev{A Graph Isomorphism Network (GIN) encodes} the resulting causal topology into a continuous soft token aligned with the language space of the Large Language Model (LLM) powering a ReAct agent\rev{, which reasons over this embedded graph to produce} structured triage reports. We evaluate \system{} on a physical, O-RAN-compliant testbed across 280 performance-degradation attack experiments. \rev{The causal ranking places the ground-truth metric within the top three candidates in 97\% of the incidents (first in 90\%). \rev{Encoding the graph as a soft token} raises the accuracy of the LLM on causal-topology queries about unseen incidents from 38\% (graph given as text) to 92\%. The agent identifies the degradation type in 90\% of the incidents and its location in 96\%.}
\end{abstract}


\maketitle

\section{Introduction}\label{sec:intro}

The transition of current mobile networks towards the Open Radio Access Network (O-RAN) architecture ~\cite{fujitsu2022vran, nttdocomo2023openran, att2023openran, gsma2021openranmou, lightreading2024orange} promises an open, virtualized, and disaggregated system with greater flexibility and cost efficiency. However, the disaggregation of traditional monolithic network functions into multiple separated components gives rise to a substantially more complex security landscape, with a greater number of interacting components from (possibly) different providers and an expanded attack surface. This complex picture is compounded by the hierarchical, one-to-many topology of the O-RAN architecture: a single O-CU serves multiple O-DUs, each of which in turn serves multiple O-RUs~\cite{ORAN_WG4_Fronthaul_CUS, oran-wg5-cp}. Attacking interfaces higher in the network’s hierarchy (e.g. the F1-C or F1-U) can degrade service for an entire cluster of cells simultaneously. 

This broader attack surface has prompted research efforts to secure exposed O-RAN interfaces~\cite{hussain20195greasoner, groen2023cost, groen2024securing, xing2024criticality, lin20255g}, develop solutions to detect attacks~\cite{scalingi2024det, wen20245g}, and incorporate resiliency mechanisms that enable the RAN to recover from platform-level failures and compute contention~\cite{garcia2021nuberu, xing2023enabling}. Among the different threats faced by O-RAN, performance-degradation attacks against O-RAN interfaces are the most prevalent threat category identified by the O-RAN Alliance Security Work Group~(WG11): 60\% of the risks catalogued in the \textit{WG11 Threat Modeling and Risk Assessment Technical Report} involve Denial of Service or performance degradation~\cite{oran-wg11-threat,baguer2024attacking}.

Despite their prevalence, investigating performance-degradation attacks is challenging because their observable consequences (e.g., high packet loss, increased latency, or higher connection drops) are indistinguishable from legitimate operational network management anomalies, including misconfigurations, software updates, or transient traffic congestion~\cite{baguer2024attacking, xing2023enabling}. While mobile networks incorporate multiple resiliency mechanisms such as redundancy and failover protocols, these defenses may be insufficient against silent compromises. When an attacker operates covertly without triggering conventional alarms, the system's inherent resilience mechanisms may inadvertently mask the attack, allowing degradation to persist undetected.

For instance, the authors in~\cite{lin20255g} demonstrate a covert man-in-the-middle (MITM) attack against the O-RAN fronthaul interface, where an attacker positioned between the O-RU and the O-DU silently degrades the control plane without triggering conventional alarms. This attack vector might be difficult to distinguish from a faulty DU, as the loss of synchronization between the O-RU and O-DU triggers similar UE disconnections~\cite{baguer2024attacking}. Notably, attacks targeting control and management interfaces (e.g., F1-C, E2, and O1) may be considerably harder to detect. Because these interfaces rely on TCP or SCTP, their inherent retransmission and connection-reset mechanisms can mask early signs of degradation. Furthermore, critical O-RAN control interfaces, such as the A1 interface, are susceptible to vulnerabilities that attackers can exploit to trigger performance-degradation attacks~\cite{thimmaraju2024security}. Such attacks can disrupt the closed-loop control mechanisms upon which O-RAN relies for intelligent network optimization. Thus, Mobile Network Operators (MNOs) are increasingly placing greater responsibility on their Network Operations Center (NOC) analysts to triage incidents under uncertainty: determining whether a performance anomaly stems from an operational fault or a potential adversarial action, locating the implicated O-RAN components, and routing the ticket to the appropriate team for deeper investigation.

Currently, to detect performance degradation events, MNOs continuously monitor a wide range of Key Performance Indicators (KPIs) (e.g., connection drop rates, cell throughput, handover success rates among others) over several-minute observation windows and typically rely on threshold-based alarms or statistical anomaly detectors~\cite{iyer2017automating, sundqvist2023robust, chawla2020interpretable, sun2024spotlight} to trigger resiliency mechanisms and open incident investigation tickets when persistent anomalies are observed. These systems flag \emph{that} a KPI deviates, but they do not answer the key question that determines the incident response workflow: \emph{was the degradation caused by an adversary who has compromised the system, by a legitimate operational fluctuation, or by a malfunction?}. This distinction is operationally decisive: it determines whether the ticket is escalated to the Security Operations Center (SOC) under the adversarial hypothesis or routed to the maintenance team as an operational fault. \rev{\system{} does not answer this question in the analyst's place; it provides the evidence the analyst needs to answer it: where the degradation enters the deployment, which metrics show it first, and which diagnostic actions can tell the hypotheses apart.}

In this paper, we present \system, \textbf{C}ausal \textbf{RE-ACT}ing agent, an agentic system designed to serve as a critical first step in O-RAN incident diagnosis. \system{} constructs metric-derived causal graphs that capture the complex inter-dependencies among O-RAN components (O-RU, O-DU, O-CU, and Radio Intelligent Controllers (RICs)). By encoding these causal graphs into continuous vector embeddings and leveraging an LLM-powered autonomous agent, \system{} generates structured preliminary diagnostic reports that rapidly narrow the scope of the incident investigation. Rather than requiring NOC analysts to manually correlate disparate system metrics and hypothesize root causes, \system{} automatically identifies the most likely suspect components and their probable failure causes, providing a focused entry point for deeper incident investigation. This initial triage significantly accelerates the incident diagnosis workflow by prioritizing which O-RAN components warrant detailed analysis, allowing analysts and downstream investigators to allocate their expertise and investigative resources efficiently during subsequent in-depth investigation phases. Our contributions are:

\begin{enumerate}
\item \textbf{Agentic triage pipeline for O-RAN}: We design \system{}, the first agentic system that supports NOC analysts in deciding how to start investigating an incident ticket in an O-RAN cellular network. \system{} rapidly produces a structured report that recommends NOC analysts a few starting investigation points and \rev{the diagnostic actions whose outcome informs the routing of the ticket. The routing decision itself remains with the analyst}. The agentic system uses a ReAct agent which iterates over a cognitive loop of interleaving reasoning and action, querying two specialized tools that turn raw O-RAN telemetry into triage-ready evidence.

\item \textbf{Weighted causal graph construction.} \rev{We adapt the Structural Agnostic Model (SAM) algorithm~\cite{kalainathan2022structural} to learn a weighted causal Directed Acyclic Graph (DAG) of an incident from O-RAN telemetry. We augment the telemetry with an artificial failure node, average the outputs of several SAM runs, project the result onto a DAG, and rank the metrics by their causal distance from the failure node. We show that both the ranking and the edges incident to the failure node are stable under telemetry noise and parameter changes.}

\item \textbf{Graph soft-prompting for LLMs.} \rev{Inspired by the GraphToken methodology~\cite{perozzi2024graphtoken}, we encode an incident's causal graph as a continuous soft token aligned with an LLM's embedding space and prepend it to a prompt that lists the graph's edges, so that a frozen Llama-3.1-8B-Instruct answers structural queries about the incident. The encoder is a Graph Isomorphism Network (GIN) that passes messages towards the failure node, trained on O-CIQA, our O-RAN Causal Inference QA dataset of 6,720 graph--question--answer triplets from 280 incidents. On incidents never seen in training, it raises accuracy on causal-topology queries from 38\% (text only) to 92\%.}

\item \textbf{Empirical validation on a physical O-RAN testbed.} We evaluate \system{} across $280$ controlled experiments spanning four O-RAN interfaces (A1, E2, F1-C, F1-U) and two impairment classes (packet loss and delay) at three (delay) or four (loss) different strength levels.

\end{enumerate}

\section{Background}\label{sec:background}
\subsection{Open RAN architecture}
\label{sec:back:oran}
The O-RAN Alliance specifies a disaggregated RAN architecture in which traditional base-station functions are distributed across three principal components. Fig.~\ref{fig:oran_arch_testbed} illustrates the O-RAN architecture and its key interfaces. The O-RAN Radio Unit (O-RU) implements the low-level physical layer (low-PHY) functions. The O-RU connects to the O-RAN Distributed Unit (O-DU) via the Open Fronthaul interface, which carries IQ samples between the low-PHY and high-PHY functions according to the O-RAN 7.2x lower-layer split. The O-DU hosts the high-PHY layer functions as well as the Medium Access Control (MAC) and Radio Link Control (RLC) layers. The O-RAN Central Unit (O-CU) supports the higher protocol layers, namely the Packet Data Convergence Protocol (PDCP), the Service Data Adaptation Protocol (SDAP), and the Radio Resource Control (RRC). The O-CU is further split into two logical entities: the O-CU Control Plane (O-CU-CP) and the O-CU User Plane (O-CU-UP). Beyond the RAN components, O-RAN introduces two Radio Intelligent Controllers (RICs) to enable data-driven, closed-loop optimization of the network. The non-Real-Time RIC (non-RT RIC) operates on timescales greater than one second and hosts rApps, which provide policy guidance, enrichment information, and machine learning (ML) model management. The near-Real-Time RIC (near-RT RIC) operates on timescales between 10\,ms and 1\,s, hosting xApps that perform fine-grained radio resource management and optimization.
\subsubsection{O-RAN Interfaces}\label{sec:back:oran:interfaces}
O-RAN standardizes the different interfaces between the components of the architecture. The interfaces relevant to this paper are the following:

\begin{itemize}
    \item \textbf{F1 Interface}: The F1 interface~\cite{oran-wg5-cp} connects the O-DU to the O-CU and is split into two components. The F1-C (F1 Control) interface links the O-DU to the O-CU-CP, carrying signaling for RRC connection management, UE context setup and release, and paging. It relies on the F1 Application Protocol (F1AP) over SCTP for reliable signaling transport. The F1-U (F1 User) interface connects the O-DU to the O-CU-UP, transporting user-plane data encapsulated with the GTP-U protocol over UDP/IP. The separation of F1-C and F1-U enables independent scaling and placement of control- and user-plane functions.
    \item \textbf{E2 Interface}: The E2 interface~\cite{oran-wg3-e2gap} connects the near-RT RIC to the RAN nodes, referred to as E2 Nodes (e.g., O-CU-CP, O-CU-UP, O-DU). It enables the near-RT RIC to collect telemetry data from the RAN and to issue control actions in near-real time. The E2 interface is the primary channel through which xApps interact with the RAN to implement closed-loop optimization.
    \item \textbf{A1 Interface}: The A1 interface~\cite{oran-wg2-a1gap} connects the non-RT RIC to the near-RT RIC. It serves three main functions: policy management, allowing the non-RT RIC to push high-level policies (e.g., QoS objectives, slicing requirements) to the near-RT RIC for enforcement; enrichment information, providing additional context such as analytics or predictions to enhance xApp decision-making; and ML model management, enabling the non-RT RIC to deploy and update ML models used by xApps in the near-RT RIC.

\end{itemize}

\section{Incident response analysis}
\label{sec:incident-response-analysis}

\subsection{Incident Life-cycle Model}
\label{subsec:incident-model}

In this section, we describe how \textit{incidents} are handled in cellular networks and frame the scope of \system{}. The O-RAN Alliance has not standardized an incident-response model; WG11's security specifications, which cover threat modeling, security requirements, protocols, and testing~\cite{oran-wg11-threat} are positioned as inputs to operator-specific security operations processes rather than as a specified response workflow. Therefore, we adopt the general-purpose NIST SP 800-61 Rev. 2~\cite{nist-sp-800-61} incident response model, instantiated for O-RAN cellular operations. The different incident response stages include the following steps:
\begin{enumerate}
    \item \textbf{Preparation}: The first stage focuses on prevention: ensuring that all components involved in the system are secure before any incident occurs. Operations teams update the different O-RAN components and configure them according to security baselines, ensuring that systems are prepared in case an adverse event occurs.
    \item \textbf{Detection and analysis}: The second step is the timely detection and accurate assessment of possible incidents. Determining whether an adverse event has occurred is an active research area in the mobile communications community~\cite{iyer2017automating, padmanabha2018mitigating, kotaru2023adapting, sun2024spotlight}. First-line anomaly detectors signal possible incidents that might be further investigated. The subsequent analysis proceeds in two stages:
    \begin{enumerate}
        \item \textbf{Triage.} The incident is classified. Two decisions are taken jointly: ($i$)~severity scoring and ($ii$) routing to the corresponding team for investigation. \rev{\system{} supports the triage decision---the most consequential human-in-the-loop step of the incident-response lifecycle---by localizing the incident and recommending further actions, but the decision itself remains with the analyst.}
        \item \textbf{Investigation and diagnosis.} The team to which the ticket was routed performs an in-depth analysis: reproducing the event, identifying the faulty or compromised component, and determining the full downstream impact.
    \end{enumerate}
    \item \textbf{Containment, Eradication, Recovery}: In this stage, the incident-handling team starts by \emph{containing} the limits of the event by isolating the affected components if possible. Next, they try to \emph{eradicate} the root cause from the system. This typically involves actions such as rolling back a recently pushed configuration or policy change, applying a vendor patch, or redeploying a degraded virtual instance from a known-good image. Finally, the handling team \emph{recovers} the system, restores full service, and verifies that the system has stabilized. 
    \item \textbf{Post-Incident Activity}: Once service is restored, the team conducts a structured review that converts the incident into organizational knowledge~\cite{nist-sp-800-61}. A lessons-learned meeting consolidates the event timeline, the diagnostic steps taken, and the remediation applied, explicitly identifying what worked, what failed, and which tooling or data sources were missing at each stage.
\end{enumerate}

\system{} supports the triage stage by producing automated hints on where the investigation should start. The deep investigation itself, as well as remediation, remains the responsibility of the corresponding investigation team; the contribution of \system{} is to narrow the search space the NOC analyst enters, not to replace the analysis performed by downstream investigation teams.

\subsection{Adversary Model}
\label{subsec:adversary-model}

The O-RAN Alliance WG11 \textit{Threat Modeling and Risk Assessment Technical Report}~\cite{oran-wg11-threat} catalogues $56$ risks across the seven O-RAN architectural domains, of which $34$ are explicitly classified as performance- degradation or denial-of-service threats. Within this scope, we consider an adversary pursuing two concurrent goals: ($i$)~\emph{degrade} the Quality of Service experienced by a subset of UEs, a single cell, or an entire cluster of cells served by a common upstream O-CU; and ($ii$)~remain \emph{indistinguishable} from benign operational faults, so that the incident is routed to the maintenance team rather than escalated to the security team. The second goal is what separates this threat class from overt denial-of-service: the adversary optimizes for \emph{triage ambiguity} rather than for maximum disruption, deliberately staying within loss and delay envelopes that are also produced by transient congestion, misconfiguration, or software regressions. A stealthy degradation absorbed into the maintenance queue is operationally more valuable than a short, loud outage escalated within minutes. The adversary operates at two levels, which may be used individually or in combination:

\begin{itemize}
\item \emph{Network-layer}: The adversary induces packet loss or adds delay to the signaling and data flows of the A1, E2, F1-C, or F1-U interfaces, using mechanisms such as traffic flooding on shared transport, or queue manipulation on intermediate switches at the transport layer~\cite{lin20255g,thimmaraju2024security}.
Attack strength ranges from subtle ($12\%$ loss or $10$\,ms added delay) to complete link failure. The adversary does not hold root access on the RIC platform or on the monitoring infrastructure and cannot decrypt interface payloads.

\item \emph{Component-layer}: The adversary controls a legitimate O-RAN component---a compromised or maliciously-authored xApp or rApp, a misbehaving O-DU image supplied by a lower-tier vendor, or a subverted CNF running on the O-Cloud---and manipulates the traffic that transits through it (by dropping, delaying, or reordering packets) without requiring access to encrypted payload contents. This level covers a significant fraction of the O-RAN Alliance WG11 high-severity threat identifiers~\cite{oran-wg11-threat} and a subset of the MITRE FiGHT 5G adversarial-technique catalogue~\cite{mitre-fight}.
\end{itemize}
\section{Problem formulation} \label{sec:problem-formulation}

\subsection{The Triage Problem} \label{subsec:the-triage-problem}

When a first-line anomaly detector flags a sustained KPI deviation in an O-RAN deployment, the NOC analyst receiving the ticket is not yet in a position to launch a deep investigation: the flagged symptom---a throughput drop, a spike in HARQ failures, a burst of UE disconnections---projects onto the radio layer regardless of where in the stack the underlying cause sits. Before any diagnostic effort is committed, the analyst must answer two coupled questions. \emph{Where} in the deployment does the investigation start: which interface, which component, and over which sub-window of the flagged observation period? And \emph{how} should the ticket be routed: to the maintenance team as an operational fault, or to the Security Operations Center (SOC) under the adversarial hypothesis of Sec.~\ref{subsec:adversary-model}? We call this combined decision the \emph{triage problem}. Its output is not a diagnosis---the deep investigation remains the investigating team's responsibility (Sec.~\ref{subsec:incident-model})---\rev{but a grounded starting point and a set of actions whose outcome informs the routing decision, which remains with the analyst. Together they narrow the search space the engineer enters.}

The triage problem is hard precisely because the symptom distribution induced by the adversary of Sec.~\ref{subsec:adversary-model} overlaps with that of a heterogeneous and open set of benign conditions that occur routinely in production O-RAN deployments~\cite{baguer2024attacking,iyer2017automating}---misconfiguration of xApp policies or slicing parameters, transient congestion on shared transport, and latent software regressions revealed by upgrades are illustrative but not exhaustive. Each deployment, vendor mix, and release cycle introduces additional failure modes whose KPI footprints partially overlap with those of the adversarial capabilities described above. \rev{\system{} therefore does not attempt to tell the adversarial and the benign hypotheses apart from the metrics alone, which would require an a priori enumeration of the look-alikes; it localizes the incident and recommends the actions whose outcome lets the analyst decide. Evaluating \system{} on benign look-alikes is left to future work (Sec.~\ref{sec:future-work}).}

\subsection{Mathematical Formulation} \label{subsec:problem-formulation}

\rev{We now state the part of the triage problem that \system{} addresses.} Let the system be observed at discrete time steps $t \in \mathbb{Z}_{\ge 0}$ with sampling period $\Delta t$, and let $\mathbf{x}[t] = (x_1[t], x_2[t], \dots, x_d[t])^\top \in \mathbb{R}^d$ denote the multivariate observation vector at step $t$, aggregating $d$ system metrics. \rev{Given an incident window $[t_s, t_e) \subset \mathbb{Z}_{\ge 0}$ of length $T_w = t_e - t_s$ flagged by a first-line anomaly detector, \system{} consumes the windowed observation $\mathcal{W} = (\mathbf{x}[t_s - T_w], \dots, \mathbf{x}[t_e - 1])$, which also covers the $T_w$ steps of normal operation that precede the incident, and produces a triage report $\mathcal{R}$.} 

\begin{equation}
     \mathcal{R} = \langle \mathcal{L}, \mathcal{C}, \mathcal{I}, \mathcal{E}, \mathcal{A} \rangle
\end{equation}

where:
\begin{itemize}
    \item $\mathcal{L}$ is a ranked list of the metrics that deviated during $[t_s, t_e)$, ordered by causal proximity to the failure;
    \item $\mathcal{C}$ and $\mathcal{I}$ are the lists of suspected components and interfaces, respectively, indicating where the analyst is advised to start the investigation;
    \item $\mathcal{E}$ is a causal evidence chain --- an ordered sequence of causally linked metric deviations connecting the observed anomalous metrics to the suspect component or interface;
    \item $\mathcal{A}$ is the set of recommended diagnostic actions whose execution helps \rev{the analyst} determine which team should receive the ticket.
\end{itemize}

A solution is acceptable when three criteria are jointly satisfied:
\begin{itemize}
    \item \textbf{Localization correctness.} The ranked metric list $\mathcal{L}$ and the suspected sets $\mathcal{C}$ and $\mathcal{I}$ agree with the ground-truth root-cause metric, component, and interface of the incident, measured through standard top-$k$ accuracy on the labeled evaluation set.
    \item \textbf{Evidence faithfulness.} Every element of $\mathcal{R}$ must be grounded either in an observation contained in the windowed observation $\mathcal{W}$ or in an edge of the causal graph constructed from it. The report must contain no hallucinated evidence --- a requirement that is non-trivial for LLM-based systems~\cite{pei2025flow, wang2024large}.
    \item \textbf{Actionability.} The recommended diagnostic actions in $\mathcal{A}$ must be executable with standard maintenance tools (e.g., terminal commands) and specific enough to name the components from $\mathcal{C}$ and interfaces from $\mathcal{I}$ they target.
\end{itemize}

\rev{We evaluate localization correctness through the top-$k$ accuracy of the ranking (Sec.~\ref{subsec:eval-causal-ranked-list}) and the identification of the attacked interface and degradation type (Sec.~\ref{subsec:eval-complete-pipeline}), and evidence faithfulness through an automatic check of every claim of the reports against the causal graph and the ranked list (Sec.~\ref{subsec:eval-final-report}). Actionability is not evaluated in this work (Sec.~\ref{sec:future-work}).}

\subsection{Challenges and limitations}
\label{subsec:approaches-limits}

Designing a triage system for O-RAN raises challenges from two distinct sources: the operational environment in which the system runs and the constraints imposed by LLMs as the underlying reasoning engine.
\begin{enumerate} 
\item[(C1)] \textbf{Telemetry volume and heterogeneity.} O-RAN generates substantially more telemetry than monolithic RAN --- logs from distributed network functions, KPI time-series, infrastructure metrics, and inter-component interaction records --- exceeding the context window of recent LLMs even when serialized aggressively.
\item[(C2)] \textbf{Causal reasoning across disaggregation.} Determining whether a degradation originates from a single compromised component or from cascading failures across O-RU, O-DU, O-CU, and the RICs requires explicit causal analysis over many fault-propagation paths, not correlation alone.
\item[(C3)] \textbf{Latency and compute budgets.} Triage sits on the critical path of the operator's Mean Time To Recovery (MTTR), so prolonged diagnosis allows stealthy attacks to persist; at the same time, operational environments impose strict compute limits that make full LLM fine-tuning on continuously generated causal graphs prohibitive.

\item[(C4)] \textbf{LLM reasoning brittleness.} Even within their context window, LLMs suffer from positional bias~\cite{liu2024lost}, hallucinate on unfamiliar domains, and require lengthy prompt engineering~\cite{wei2022chain, zhang2023meta} that aggravates context pressure. They also struggle with telecommunications vocabulary and structured modalities such as time series and packet captures~\cite{Liu_Zhang_Qian_Ma_Qin_Bansal_Lin_Rajmohan_Zhang_2024, Wu_Wang_Qiao_Wang_Jiang_Cui_Wang_2024}.

\item[(C5)] \textbf{Iterative summarization is not a fix.} Repeatedly summarizing telemetry to fit the context window introduces generative randomness that can drop critical signals~\cite{wang2024large}, and large telemetry volumes overwhelm such pipelines regardless.
\end{enumerate}

\system{} addresses these challenges through two design decisions. First, rather than prompting the LLM with raw telemetry, we construct a compact causal representation of the failure that surfaces likely root causes within a small context window, addressing C1--C2 and C4--C5 within the latency budget of C3. Second, we ground the agent's reasoning through a specialized graph encoder fine-tuned on the graph modality, training only very few parameters compared to the total LLM's parameters and thereby satisfying the compute side of C3 alongside the brittleness in C4.

\section{\system{}}
\label{sec:creact} 

\subsection{System Model}
\label{subsec:system-model}
\system{} is designed as a post-detection triage stage that plugs into the existing observability stack of an O-RAN operator. It does not replace any first-line anomaly detector, nor the in-depth investigation performed by a team of operations or security engineers; rather, it sits between them and converts a set of persistent incident signals (i.e. anomalous Key Performance Indicators (KPIs)) into an initial triage report that aids the analyst in deciding where to start investigating and helps in classifying and routing incident tickets. Consistent with the incident model of Sec.~\ref{subsec:incident-model}, the passive observability pipeline that supplies \system's inputs, together with \system{} itself (its agentic tools and reasoning agent), forms the Trusted Computing Base (TCB). We assume the adversary cannot tamper with TCB outputs. The O-RAN components under analysis — xApps, rApps, O-DU, O-CU, RIC platform services, and their hosting O-Cloud nodes — lie outside this envelope and may be faulty, misconfigured, or compromised.

\system{} is triggered when an upstream anomaly detector flags a sustained KPI deviation within a bounded observation window $[t_s, t_e]$. We assume that commercial mobile networks include resiliency mechanisms that allow the system to recover from many incidents autonomously~\cite{xing2023enabling, garcia2021nuberu}. Thus, \system{} uses its agentic tools to process different metric representations during the incident window and produce the report $\mathcal{R}$ within the triage-stage latency budget (typically tens of seconds to a few minutes; measured in Sec.~\ref{subsec:eval-execution-time}), providing NOC analysts with a grounded entry point for further investigation.

\begin{figure*}[t]
  \centering
  \includegraphics[width=\linewidth]{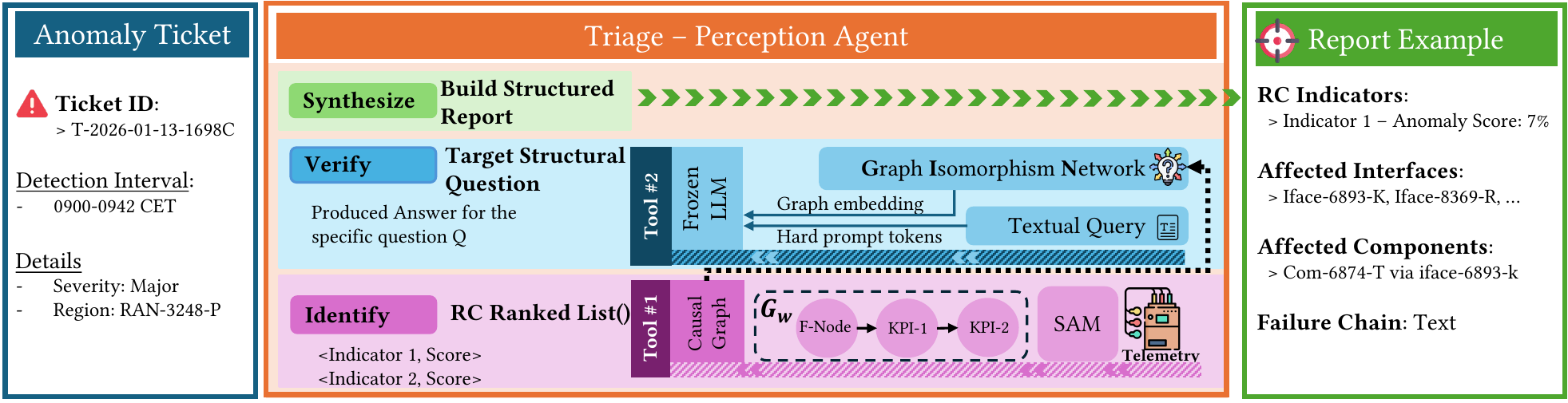}
    \caption{\system{} architecture. Stage~1 constructs a weighted causal graph $\mathcal{G}_W$ over interface metrics. Stage~2 compresses $\mathcal{G}_W$ into a continuous soft token aligned with the LLM embedding space to allow the LLM to answer structural queries. Stage~3 is a ReAct-based agent that iteratively queries both tools to produce the triage report $\mathcal{R}$.}
  \label{fig:architecture}
\end{figure*}

\subsection{Agentic design}
\label{subsec:agentic-design}
Generating the triage report $\mathcal{R}$ requires three different kinds of reasoning: \emph{quantitative ranking} of the metrics that are causally closest to the failure; \emph{structural reasoning} over a causal graph of metrics identifying interfaces, components, and propagation paths; and \emph{linguistic synthesis} that composes the first two into an auditable report populated with O-RAN domain knowledge. \system{} adopts an agentic architecture~\cite{yao2023react} that separates the three concerns. A ReAct-based autonomous agent, implemented as a LangGraph~\cite{langgraph} state machine, provides the linguistic synthesis and the high-level control flow; two specialized tools, exposed to the agent through LangGraph's tool-calling interface, provide the quantitative and structural primitives on which the agent reasons. The two tools correspond to the two sources of uncertainty in the triage problem: which metric, among the tens of correlated ones in the incident observation window, \rev{is most directly affected by the incident} (Sec.~\ref{subsec:causal-graph}); and which interface or component the implicated metrics belong to in the deployment topology (Sec.~\ref{subsec:soft-prompt}). The system prompt, reported in Appendix~\ref{subsec:appendix-input-output-agent}, defines its persona as an expert O-RAN Site Reliability Engineer. The agentic workflow follows a three-phase triage protocol. 
\begin{itemize}
\item \textbf{Identify}: The agent uses the \texttt{ranked\_list\_causes} tool to retrieve a ranked list of metrics causally closest to the failure node \fnode{}, an artificial node we add to the graph to represent the incident itself.  
This phase populates $\mathcal{R}$ with an initial set of suspect components and interfaces, scoping the search space for downstream severity scoring and ticket routing.

\item \textbf{Verify}: For each of the top candidates, the agent issues targeted structural queries using the tool \texttt{query\_graph}. 
For instance, the ReAct agent can ask ``is the E2 interface affected?'' or ``how many F1-c KPIs are directly connected to the root incident node (\fnode{})?'' to confirm or refute the incident propagation path implicit in the ranked list. The answers returned by the tool, populate $\mathcal{C}$ and $\mathcal{I}$ and anchor the edges of the evidence chain $\mathcal{E}$.
\item \textbf{Synthesize}: The agent composes a coherent failure chain that connects the root-cause metric in $\mathcal{R}$ to the affected interfaces in $\mathcal{I}$ and components in $\mathcal{C}$, and emits the structured report $\mathcal{R}$. The synthesis step also derives the recommended diagnostic actions $\mathcal{A}$, \rev{each naming the components of $\mathcal{C}$ and the interfaces of $\mathcal{I}$ it applies to and expressed in the operator's tooling vocabulary (e.g.\ ``perform a network trace on the E2 interface to identify where the latency is being introduced''), so that the outcome of executing $\mathcal{A}$ lets the analyst decide the final routing of the ticket.}
\end{itemize}
 
The agent runs a standard \texttt{call\_model}--\texttt{tools} loop that terminates when the LLM decides no further tool invocation adds evidence, subject to a maximum-step budget of $K=40$ loops. A final \texttt{generate\_structured\_response} step issues one additional LLM call that emits $\mathcal{R}$ in a machine-parseable schema alongside the natural-language rationale, finalizing the overall process.

\subsection{Tool~\#1: Causal Ranked List} 
\label{subsec:causal-graph}

The agentic tool \texttt{ranked\_list\_causes} supplies to our system a quantitative ranking of the metrics that are most closely related to an incident. This tool is used during the \emph{Identify} Phase. The tool operates on the multivariate observation $\mathbf{x}[t]$ introduced in Sec.~\ref{subsec:problem-formulation}, taken over the contextual window $C_W$
\begin{align}
C_W = [t_s - T_w,\, t_e), \qquad T_w = t_e - t_s
\label{eq:context-window}
\end{align}
which extends $T_w$ seconds before the incident window flagged by a first-line anomaly detector. This symmetric extension exposes both the normal and incident regimes of the system to the causal discovery algorithm (see Sec.~\ref{subsubsec:sam}), allowing it to identify the metrics whose joint behavior changes between the two regimes and so distinguishing genuine causal links from coincidental correlations present in the baseline. Using the contextual window, the tool learns a Weighted Directed Acyclic Graph

\begin{align}
\mathcal{G}_W = \bigl(V \cup \{\mathcal{F}, \mathcal{U}\},\; \mathbf{E}\bigr), \qquad \mathbf{E} \in [0,1]^{n \times n}
\label{eq:graph}
\end{align}

where $V$ is the set of nodes representing each metric variable; $\mathcal{F}$ and $\mathcal{U}$ are special root nodes encoding the incident signal and UE-population dynamics, called \emph{Failure Node} and \emph{UEs Node} (see Sec.~\ref{subsubsec:build_node_set}); and $\mathbf{E}$ is the weighted adjacency matrix of the graph, with $\mathbf{E}_{ij} = 0$ encoding the absence of an edge from node $i$ to $j$ and $\mathbf{E}_{ij} > 0$ encodes a directed edge from node $i$ to $j$ of causal-strength $\mathbf{E}_{ij}$.

\subsubsection{Building the Node Set}
\label{subsubsec:build_node_set}

The \texttt{ranked\_list\_causes} tool builds the node set $V$ from the multivariate observation vector $\mathbf{x}[t]$ introduced in Sec.~\ref{subsec:problem-formulation}. Each time series $x_i[t]$, $t \in C_W$ becomes a node $v_i \in V$. 
Two additional time series are appended to $\mathbf{x}[t]$ and added as nodes of $\mathcal{G}_W$:  
\begin{itemize}
\item \textbf{Failure Node $\mathcal{F}$:} a binary indicator time series set to $0$ over $[t_s - T_w,\, t_s)$ and $1$ over $[t_s,\, t_e)$, encoding the incident signal produced by the first-line anomaly detector without any manual labeling at inference time. \rev{The failure node follows the failure-indicator node of RCD~\cite{ikram2022root}.}

\item \textbf{UEs Node $\mathcal{U}$:} an integer-valued time series counting the registered UEs in the network at each sample over the same window, derived from the access-and-mobility counters exposed by the O-DU.
\end{itemize} 

The graph has $n = d + 2$ nodes: $d$ metric nodes plus $\mathcal{F}$ and $\mathcal{U}$. Before learning the causal graph (see Sec.~\ref{subsubsec:sam}), each time series in $\mathbf{x}[t]$ (including $\mathcal{F}$ and $\mathcal{U}$) \rev{is standardized to zero mean and unit variance over $C_W$ (a constant series is left at zero)} 
to remove unit-of-measurement effects and place all metrics on a comparable scale.

\subsubsection{Structural Agnostic Model (SAM) Causal Discovery}
\label{subsubsec:sam}

To uncover the causal structure among the variables in $V\cup\{\mathcal{F},\mathcal{U}\}$, we adopt the Structural Agnostic Modeling (SAM) algorithm~\cite{kalainathan2022structural}. Unlike constraint-based discovery methods that rely on pairwise conditional-independence tests~\cite{spirtes2000causation,ikram2022root}, SAM formulates causal discovery as a continuous optimization problem and uses Generative Adversarial Networks to model both conditional independencies and distributional asymmetries jointly. \rev{This formulation avoids pairwise conditional-independence tests, which become unreliable as the number of correlated variables grows.}
For each node $v_j\in V \cup \{\mathcal{F},\mathcal{U}\}$ associated with a time series $x_j[t]$, SAM instantiates a conditional generative network $\hat{f}_j$ implemented as a $H$-hidden-layer neural network that reconstructs $x_j[t]$ from the other variables as 

\begin{align}
\hat{x}_j[t] \;=\; \hat{f}_j\!\left(\mathbf{x}_{-j}[t],\; a_j,\; \theta_j\right), \quad t \in C_W
\label{eq:sam-generator}
\end{align}

where $\mathbf{x}_{-j}[t]$ is the set of all time series except $x_j[t]$, $\theta_j$ are the conditional generative network parameters (i.e., all weight matrices and bias vectors), and $a_j \in [0,1]^n$ is a learned structural gate vector --- driven toward $\{0, 1\}$ by the sparsity penalty introduced below --- that acts as the $j$-th column of the adjacency matrix and decides which variables effectively act as parents of $v_j$. The structural gates of all variables, stacked column-wise, form the gate matrix $A = [a_1, \dots, a_n]$. The gate value $A_{ij} = (a_j)_i$ has a direct causal interpretation: it is non-zero only when including $x_i[t]$ as a predictor for $x_j[t]$ genuinely improves the reconstruction; thus, $x_i[t]$ is causally related to $x_j[t]$.

SAM is trained as a Generative Adversarial Network (GAN)~\cite{ goodfellow2014generative}. The per-variable generative networks $\hat{f}_j$ act as \emph{generators} that reconstruct each metric from its candidate parents, while a single neural \emph{discriminator} with parameters $\omega$ is trained to tell apart the real samples drawn from $\mathbf{x}[t]$ from the synthetic samples the generators produce. Generators and discriminator are trained against each other. The generators learn to fool the discriminator while the discriminator learns to spot them. The gate matrix $A$, the network weights $\theta=\{\theta_j \}_{j=1}^n$ and the discriminator weights $\omega$ are learned jointly by minimizing 

\begin{equation}
\begin{split}
\mathcal{L}_{\text{SAM}} = \min_{A, \theta} \max_{\omega} \Bigl(
  &\mathcal{L}_{\text{fit}} + \lambda_S \sum_{i,j} a_{i,j} \\
  &+ \lambda_F \sum_j \lVert \theta_j \rVert_F
   + \lambda_D \sum_{k=1}^{n} \tfrac{\mathrm{tr}(A^k)}{k!} \Bigr),
\end{split}
\label{eq:sam-loss}
\end{equation}

where the min--max structure reflects the adversarial game between the generators (parameterized by $A$ and $\theta$) and the discriminator (parameterized by $\omega$). The first term, $\mathcal{L}_{\text{fit}}$, is the adversarial fitting loss that drives each generator $\hat{f}_j$ to produce reconstructions of $\hat{x}_j[t]$ that the discriminator cannot distinguish from the true data, given its currently active parents set by matrix $A$. The second term penalizes the $\ell_1$ norm of $A$ and enforces the \emph{sparsity} prior that a metric depends on a small number of direct causes rather than on all the others~\cite{hastie2015statistical}. The third term regularizes the Frobenius norm of the generator weights, controlling the \emph{functional complexity} of each $\hat{f}_j$ and avoiding overfitting on the small contextual window. The fourth term measures the weight of directed cycles of every length $k$ in $A$ and is zero only for acyclic matrices; minimizing it enforces the absence of cycles in $A$. The hyperparameters $\lambda_S$, $\lambda_F$, and $\lambda_D$ control the trade-off between fit, parsimony, regularization, and acyclicity; their values are reported in Appendix~\ref{appendix:sam_hyperparameters}. Nodes $\mathcal{F}$ and $\mathcal{U}$ act as exogenous drivers of the system rather than consequences of it. We enforce this by masking the columns of $A$ corresponding to $\mathcal{F}$ and $\mathcal{U}$ throughout training, fixing $a_{i, \mathcal{F}} = a_{i, \mathcal{U}} = 0$ for all $i$ and so placing both nodes as roots in the recovered DAG. 

\rev{To reduce the variance of a single SAM run, we execute $M=16$ instances with distinct random seeds and average their learned gate matrices into $\bar{A} = \frac{1}{M} \sum_{m=1}^{M} A^{(m)}$. Averaging does not preserve acyclicity, because different runs can orient an edge in opposite directions. We therefore project $\bar{A}$ onto a DAG in three steps: ($i$) keep the edges with $\bar{A}_{ij} \ge 0.5$; ($ii$) for every pair of opposite edges, keep the stronger one; ($iii$) while a directed cycle remains, remove its weakest edge. The result is the weighted DAG $\mathcal{G}_W$, with $\mathbf{E}_{ij} = \bar{A}_{ij}$ on the kept edges and $\mathbf{E}_{ij} = 0$ elsewhere.}

\subsubsection{Ranking}
\label{subsubsec:ranking}
\rev{The \texttt{ranked\_list\_causes} tool derives a ranked list $\mathcal{L}$ of all metrics, ordered by their causal proximity to $\mathcal{F}$. Each directed edge $(v_i, v_j)$ of the averaged gate matrix carries a causal-strength weight $\bar{A}_{ij} \in [0, 1]$, where larger values indicate stronger evidence that $v_j$ is causally driven by $v_i$. We define the \emph{distance} of the edge $(v_i, v_j)$ as}
\begin{align}
\delta(v_i, v_j) \;=\; 1 - \bar{A}_{ij}.
\label{eq:edge-distance}
\end{align}
\rev{The distance of metric $v_j$ from $\mathcal{F}$ is the length of the shortest path from $\mathcal{F}$ to $v_j$:}
\begin{align}
d_{\mathcal{F}}(v_j) \;=\; \min_{\pi:\, \mathcal{F} \rightsquigarrow v_j} \; \sum_{(v_a, v_b) \in \pi} \delta(v_a, v_b).
\label{eq:distance-f}
\end{align}
\rev{A metric with a strong direct edge from $\mathcal{F}$ has $d_{\mathcal{F}}$ close to $0$; a metric reached only through weak or long paths has a large $d_{\mathcal{F}}$. The shortest paths are computed on $\bar{A}$ rather than on the thresholded $\mathcal{G}_W$, so every metric has a finite distance and the ranking does not depend on the edge threshold. Sorting the metrics in ascending order of $d_{\mathcal{F}}$ yields the ranked list $\mathcal{L}$ that the agent retrieves through the \texttt{ranked\_list\_causes} tool call (Phase~1 of the diagnostic protocol, Sec.~\ref{subsec:agentic-design}). The node $\mathcal{U}$ takes no part in the ranking; it is part of the graph so that SAM can attribute changes caused by UE attachment to $\mathcal{U}$ rather than to $\mathcal{F}$.}

\rev{The tool returns the complete ranked list $\mathcal{L}$ to the agent: for each of the $d$ metrics, its name, its distance $d_{\mathcal{F}}(v)$ and its rank. Metric names carry the O-RAN interface they are measured on (e.g.\ \texttt{e2\_du\_latency\_mean\_ul\_ms} belongs to E2), so the agent can map each metric to an interface without further input. Because every gate is below one, $\delta > 0$, so a metric reached from $\mathcal{F}$ through an intermediate metric $u$ is always farther from $\mathcal{F}$ than $u$. The top-ranked metric $r$ is therefore always reached by its own edge from $\mathcal{F}$, with $d_{\mathcal{F}}(r) = 1 - \bar{A}_{\mathcal{F}r}$, the strongest gate leaving $\mathcal{F}$: the metric through which the incident shows first. This edge belongs to $\mathcal{G}_W$ whenever that gate reaches the threshold. Otherwise, no metric is linked to the incident with sufficient confidence. Lower-ranked metrics may be reached through intermediate metrics. The ranked list does not tell which of these cases applies, so the agent establishes the position of each candidate in $\mathcal{G}_W$ through the structural queries of Tool~\#2 (Sec.~\ref{subsec:soft-prompt}). Tool~\#1 also outputs $\mathcal{G}_W$ and the onset $\tau(r)$ of the top-ranked metric, the first second after $t_s$ at which it deviates from its pre-incident mean by more than $2\sigma$. Neither is passed to the agent, but both let the analyst check the report against the graph and the measurements.}

\begin{figure*}[ht!]
    \centering
    \includegraphics[width=\linewidth]{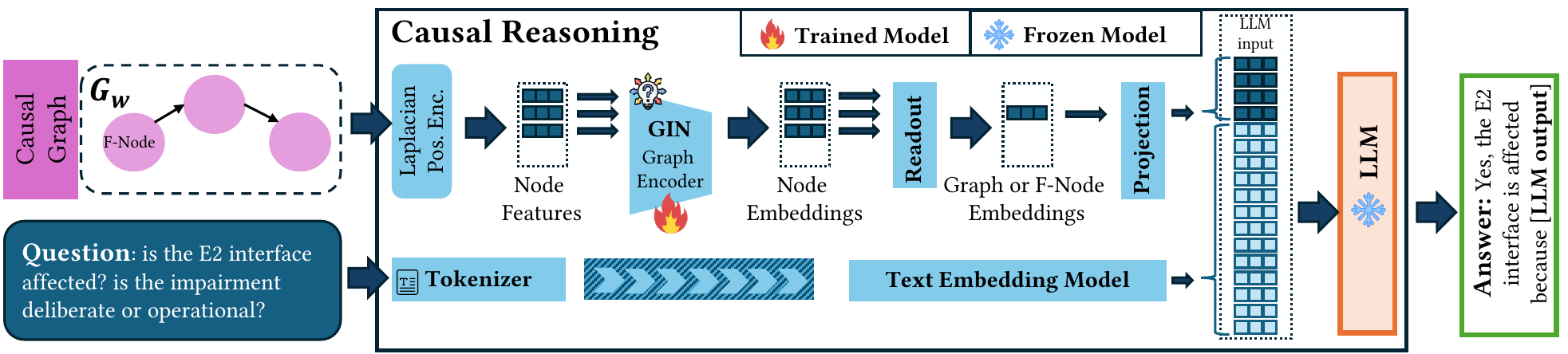}
    \caption{Graph Soft-Prompting architecture. By updating only the GIN encoder weights while keeping the LLM completely frozen, it compresses causal graphs into continuous soft tokens with minimal computational overhead.}
    \label{fig:graph_soft_prompting_arch}
\end{figure*}

\subsection{Tool~\#2: Causal Reasoning via Graph Soft-Prompting}
\label{subsec:soft-prompt}

In the \emph{Verify} phase of \system{}'s agentic design, the agent reasons about the structure of the causal graph $\mathcal{G}_W$ computed by Tool~\#1 (\texttt{ranked\_list\_causes}, see Sec.~\ref{subsec:causal-graph}), to decide which O-RAN interfaces or components were the first to deviate from normal operation, and whether any unexpected behavior propagated to other components or interfaces. The output of this reasoning step populates the suspect sets $\mathcal{C}$ and $\mathcal{I}$ in the triage report~$\mathcal{R}$. The agentic tool \texttt{query\_graph} provides this structural-reasoning primitive. Given a natural-language query $Q$ about $\mathcal{G}_W$ generated by the agent (e.g. ``Is the F1-C interface affected?'', ``Which F1-U Key Performance Metrics (KPIs) are directly connected to $\mathcal{F}$?'', ``\rev{How many A1 interface KPIs are affected by the anomaly?}''), the tool encodes $\mathcal{G}_W$ as a single continuous \emph{soft token} aligned with the embedding space of a frozen large language model (LLM), prepends that token to $Q$, and returns the LLM's textual answer to the agent. \rev{We adopt a soft-prompting methodology similar to GraphToken~\cite{perozzi2024graphtoken}: the soft token carries the structure of $\mathcal{G}_W$ into the LLM's attention layers, and the prompt also lists the nodes and directed edges of $\mathcal{G}_W$ with their metric names, since the soft token encodes structure but not which metric each node stands for. Fig.~\ref{fig:graph_soft_prompting_arch} shows the four-stage pipeline of \system{}'s graph encoder architecture.}

\subsubsection{Graph Encoding}
\label{subsubsec:encoder}
\rev{To encode $\mathcal{G}_W$ into a soft token, we first represent its nodes as $4$-dimensional vectors using a Laplacian Positional Encoding (LPE)~\cite{dwivedi2023benchmarking}. Let $E$ be the weighted adjacency matrix of the DAG of Sec.~\ref{subsec:causal-graph}, with $E_{ij}$ the gate of edge $i \rightarrow j$. The LPE describes where a node sits in the graph, regardless of edge direction, so it is computed on the symmetrized matrix $W = E + E^\top$, restricted to the set $V'$ of nodes with at least one edge. On $V'$ every degree $D_{ii} = \sum_j W_{ij}$ is positive, and the normalized Laplacian
\begin{align}
L = I - D^{-1/2} W D^{-1/2}
\label{eq:laplacian}
\end{align}
is symmetric and positive semi-definite, with real eigenvalues in $[0, 2]$. The LPE of a node consists of its entries in the eigenvectors of the four smallest strictly positive eigenvalues of $L$ (zero-padded if fewer exist). The eigenvectors of eigenvalue zero, which only mark connected components, are skipped, and nodes without edges receive a zero encoding. Eigenvectors are defined only up to sign, and up to rotation within a repeated eigenvalue, which occurs for structurally identical components such as several two-node throughput pairs. During training we therefore draw a random orthogonal rotation within every group of equal eigenvalues (a random sign flip for a simple eigenvalue), so that the encoder does not depend on this arbitrary choice, \rev{as in \cite{dwivedi2023benchmarking}}. Stacking the encodings row by row yields a matrix of size $(d+2) \times 4$, where $d$ is the number of metrics.}

Following the LPE coordinate assignment, the node features of $\mathcal{G}_W$ are projected into a continuous latent space using a Graph Isomorphism Network (GIN)~\cite{xu2019powerful}. GINs are a class of Graph Neural Networks (GNNs) theoretically proven to maximize structural distinguishability, ensuring that non-isomorphic local structures map to distinct representations. The resulting embeddings capture both the global spatial coordinates of a node within $\mathcal{G}_W$ and its specific local causal topology. \rev{We employ a stack of three GIN layers with a latent dimension of $128$. Edge direction, and with it the causal information, enters through message passing: the GIN runs on the \emph{reversed} graph, so that every node aggregates the representations of its children, scaled by the gates of $E$, and combines this aggregate with its own state:
\begin{align}
h_v^{(l+1)} = \mathrm{ReLU}\Big(\Theta^{(l)}\big[(1+\epsilon)\, h_v^{(l)} + \textstyle\sum_{u:\, v \rightarrow u} E_{vu}\, h_u^{(l)}\big] + b^{(l)}\Big),
\label{eq:gin}
\end{align}
with $h_v^{(0)}$ the LPE, $\epsilon = 0.1$, and the weights $\Theta^{(l)}$ and biases $b^{(l)}$ learned for each layer $l \in \{0,1,2\}$. After the three layers, the embedding $h^{(3)}_v \in \mathbb{R}^{128}$ of node $v$ summarizes its descendants within three causal links; in particular, the failure node $\mathcal{F}$, the root of $\mathcal{G}_W$, aggregates the metrics that the incident reaches within three hops. Stacking the per-node embeddings row by row yields $H^{(3)} \in \mathbb{R}^{(d+2) \times 128}$.}

To convert $H^{(3)}$ into a single soft token for LLM consumption, we execute a \emph{readout} step that aggregates the rows of $H^{(3)}$ into one fixed-size vector. The choice of aggregation strategy is critical: distinct categories of structural queries depend on different topological features of the graph. The natural-language queries submitted to Tool~\#2 fall into two structural categories:
\begin{itemize}
    \item \emph{Graph-level queries} (e.g., ``Is the F1-c interface affected?'') require a complete representation of the graph, since the agent must reason about propagation paths and cascading effects that may impact interfaces not directly adjacent to $\mathcal{F}$.
    \item \emph{Node-level queries} (e.g., ``\rev{List the F1-c interface KPI nodes that are directly connected to the F-node.}'') interrogate the immediate topological neighborhood of $\mathcal{F}$.
\end{itemize}
A single one-size-fits-all aggregation discards information that one query type needs but another can ignore; we therefore use two different readout strategies, dispatched by query type.

For \emph{graph-level} queries, we apply mean pooling across all node embeddings to synthesize a global graph vector:
\begin{align}
h_G = \frac{1}{N} \sum_{i=1}^{N} h_i^{(3)},
\label{eq:readout-graph}
\end{align}
where $N = d + 2$ is the total number of nodes in $\mathcal{G}_W$. This collapses the $N \times 128$ embedding matrix $H^{(3)}$ into a single $128$-dimensional vector via element-wise arithmetic averaging.

For \emph{node-level} queries, we extract the F-Node embedding $h_{\mathcal{F}}^{(3)}$ directly. Because the GIN encodes each node with three hops, this single vector encapsulates the localized topology and causal dynamics within a $3$-hop radius of $\mathcal{F}$, providing the LLM with a concentrated structural fingerprint of the fault's immediate vicinity.

While both readout strategies share the GIN architecture, they are optimized as distinct model checkpoints. At inference time, the agent's natural-language query is classified into one of the two categories by a lightweight rule-based classifier on the query template, and the tool dispatches the forward pass through the corresponding checkpoint. The selected readout vector, denoted generically as $h \in \mathbb{R}^{128}$, is transformed by a linear projection layer $g_\phi: \mathbb{R}^{128} \to \mathbb{R}^{4096}$ into a continuous soft token aligned with the native embedding space of a Llama-3.1-8B-Instruct model~\cite{grattafiori2024llama3}. The final input prompt sequence $P$ is constructed via concatenation:
\begin{align}
P = g_\phi(h) \mathbin{\Vert} t(Q),
\label{eq:prompt}
\end{align}
where $g_\phi(h)$ is the projected graph embedding, \rev{$t(Q)$ is the token embedding sequence of the text prompt, i.e., the list of nodes and directed edges of $\mathcal{G}_W$ followed by the natural-language query $Q$}, and $\mathbin{\Vert}$ denotes the concatenation operator.

\subsubsection{Soft-Token Encoding}
\label{subsubsec:training}

Tool~\#2 must be trained so that the GIN learns to encode causal graphs into soft tokens that steer the frozen LLM toward correct answers. Each training example is a triplet $(\mathcal{G}_W, Q, y)$ where $y$ is the ground-truth answer derived from the labeled incidents generated in our O-RAN testbed (see Sec.~\ref{sec:eval}). After concatenating the graph soft token with the tokenized query and passing the result to the frozen LLM, the model emits a probability distribution over its output vocabulary. The training signal is the negative log-likelihood of the correct answer \rev{computed over the answer tokens, including the end-of-sequence token, on the training pairs only}:
\begin{align}
\mathcal{L}(\phi, \theta_{\text{GIN}}) \;=\; -\sum_{i=1}^{n} \log p_{\phi, \theta_{\text{GIN}}}\!\left(y_i \mid P_i\right),
\label{eq:soft-prompt-loss}
\end{align}
where $\phi$ and $\theta_{\text{GIN}}$ encapsulate the learnable parameters of the linear projection layer and the GIN encoder respectively, $n$ is the batch size, $P_i$ is the input prompt sequence of Eq.~\ref{eq:prompt} for instance $i$, and $p_{\phi, \theta_{\text{GIN}}}(y_i \mid P_i)$ is the LLM's predicted probability of emitting the ground-truth answer $y_i$ given $P_i$. Since the LLM weights remain frozen, gradients update only $\phi$ and $\theta_{\text{GIN}}$, refining the soft tokens to concentrate the LLM's probability mass on the correct text output.
\section{Evaluation} \label{sec:eval}

\begin{figure}[t!]
  \centering
  \includegraphics[width=0.80\linewidth]{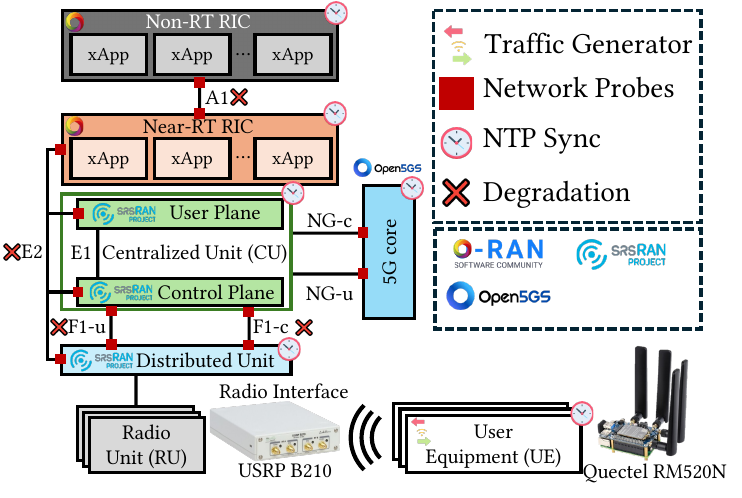}
  \caption{Experimental testbed used for the evaluation.}
  \label{fig:oran_arch_testbed}
  \vspace{-11mm}
\end{figure} 

\subsection{Experimental Setup}
\label{subsec:experimental-setup}

To evaluate \system{}, we built the experimental O-RAN testbed shown in Fig.~\ref{fig:oran_arch_testbed}. We decided to physically distribute the testbed components across three dedicated host machines interconnected via a high-speed LAN, keeping all hosts' clocks synchronized. The primary host (AMD Ryzen R9 7950X, 64GB RAM, Ubuntu 22.04 LTS) allocates the O-RAN components O-RU, O-DU, and O-CU, the O-RAN RICs, and 5G Core. To satisfy the hard real-time requirements of the O-DU and prevent noisy-neighbor interference during long-running experiments, all O-RAN components are containerized using Docker and assigned dedicated CPU cores. A second host (Intel i9-13900H, 32GB RAM) manages the user equipment (UE), and the third host is dedicated to the observability stack.
The software stack relies on open-source solutions fully aligned with 3GPP and O-RAN Alliance standards: the O-RAN Software Community ~\cite{oran_sc} (for the Non-RT and near-RT RICs), srsRAN ~\cite{srsran} (for the O-RU, O-DU, and O-CU), and Open5GS~\cite{open5gs} (for the Core Network). The physical radio link is established using an Ettus USRP B210 Software-Defined Radio (SDR) acting as the O-RU radio head and two Quectel RM520N-GL serving as standard-compliant UEs.
\rev{The agent runs on a server with two NVIDIA GH200 GPUs: its language model, Gemma~4 31B~\cite{gemma4, team2024gemma} quantized to 4 bits (GGUF Q4\_K\_M) and served with Ollama, uses one GPU, and Tool~\#2, with Llama-3.1-8B-Instruct in bfloat16, the other. Every call of both language models uses greedy decoding (temperature~0), and the agent uses a fixed seed, so the evaluation is repeatable. Tool~\#1 also runs on the same host; the 16 SAM runs of an incident train concurrently under NVIDIA's Multi-Process Service (MPS).}

\subsubsection{Data Observability}
To capture a comprehensive and highly granular view of the system's state, we continuously retrieve interface metrics at each O-RAN component \rev{at both ends of each interface: F1-C and F1-U at the O-CU and the O-DU, E2 at the O-DU and the near-RT RIC, and A1 at the non-RT RIC and the near-RT RIC. For each interface and direction, the metrics are the mean and 90th-percentile one-way latency, the packet loss, the number of retransmissions and the throughput. Packets are matched between the two ends by their protocol fields (SCTP TSNs and chunk fields, TCP sequence numbers, GTP-U IP identifier and payload hash). Latency is the one-way delay of matched packets, and a packet with no match within 10\,s counts as lost. Packet loss is computed over the last 10\,s \rev{window}. Latency below 1\,ms is set to zero: outside a delay attack it never exceeds 0.5\,ms (99\% of the values are below 0.02\,ms), while the smallest injected delay is 10\,ms. Seconds without packets keep the last measured latency.}

\begin{figure}[t!]
  \centering
  \includegraphics[width=1\linewidth]{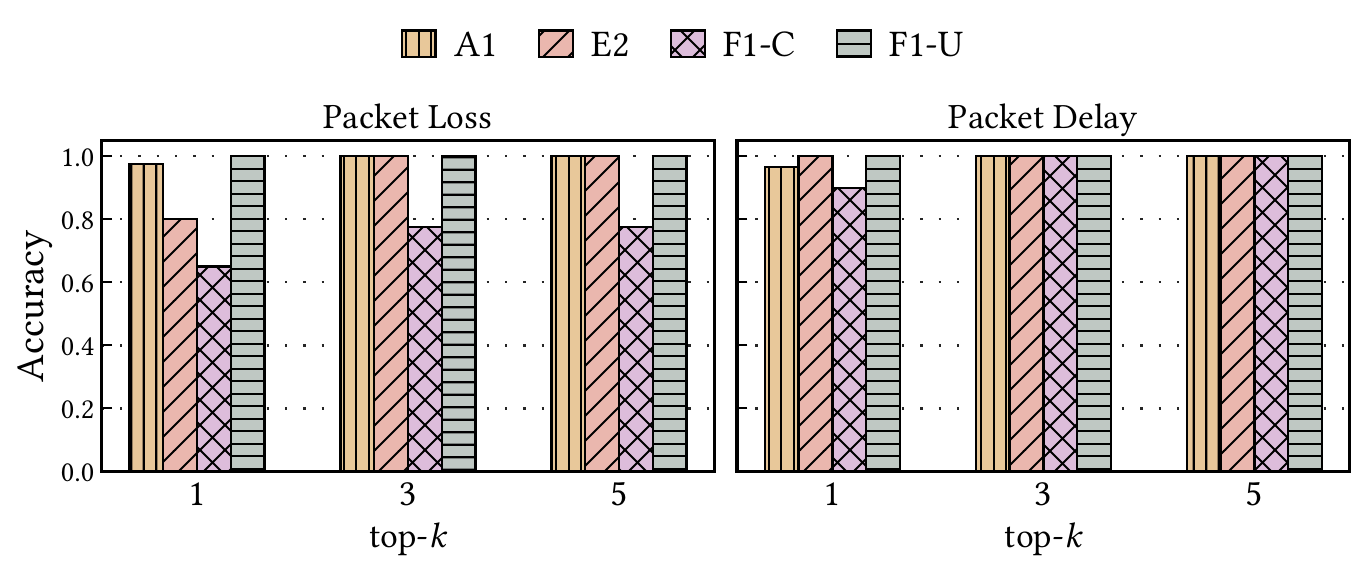}
  \caption{\rev{Top-$k$ accuracy of the \texttt{ranked\_list\_causes} tool per attacked interface and attack type. Each group covers 30 delay or 40 loss incidents.}}
  \label{fig:root-cause-ranking-evaluation_k}
\end{figure}

\subsubsection{Traffic Generation and Adversarial Emulation} \rev{To move beyond synthetic traffic patterns, the testbed generates downlink user traffic with MGEN~\cite{mgen}, replaying traces of a live commercial LTE cell in Spain, recorded with the FALCON control-channel monitor~\cite{Falkenberg2019a}. \rev{This means that each UE attached during the experiment mimics the traffic pattern of a random flow recorded with FALCON}. To emulate the adversarial capabilities described in Sec.~\ref {subsec:adversary-model}}, where an attacker silently degrades O-RAN interfaces, we inject different levels of packet loss and packet delay into F1-U, F1-C, A1, and E2 interfaces. This effectively reproduces the operational footprint of a performance-degradation attack.

\subsubsection{\rev{Dataset}}
\label{subsubsec:dataset}
\rev{We recorded 280 attack runs: for each of the four interfaces (A1 downlink, E2 uplink, F1-C downlink, F1-U downlink), packet delay at three strengths and packet loss at four strengths, with ten runs per configuration, i.e., 30 delay and 40 loss runs per interface. Packet loss is 12--30\% (low), 40--60\% (medium), 70--89\% (high) or 100\% (link down); delay is 10--94\,ms (low), 111--992\,ms (medium) or 1.07--4.91\,s (high). In each run the attack starts 80--120\,s after the start of the run and lasts 82--159\,s, and a second UE attaches about 150\,s after the start, while the attack is active. The incident window $[t_s, t_e)$ is the attack window recorded by the experiment, i.e., we assume a first-line detector that flags the attack exactly.}

\subsection{\system{}'s Tool~\#1 Evaluation}
\label{subsec:eval-causal-ranked-list}
\subsubsection{\rev{Metrics ranking}}

\rev{We first evaluate how well the \texttt{ranked\_list\_causes} tool ranks the ground-truth metrics, i.e., the metrics that the injected attack changes directly: the packet loss of the attacked direction for loss attacks, and the mean or 90th-percentile latency of that direction for delay attacks. An incident counts as correct at $k$ when a ground-truth metric is among the first $k$ metrics of the ranked list.}

\rev{Figure~\ref{fig:root-cause-ranking-evaluation_k} reports the top-$k$ accuracy for each attacked interface and attack type. Over the 280 incidents, Tool~\#1 ranks the ground-truth metric first in 90.4\% of them, within the top three in 96.8\% and within the top five in 96.8\%. Delay is ranked within the top three in all 120 delay incidents, and first in all E2 and F1-U delay incidents, in 29 of 30 on A1 and in 27 of 30 on F1-C. Loss is ranked within the top three in all A1, E2 and F1-U loss incidents. On E2, loss is first in 32 of 40 incidents; in the other eight, the first metric is the E2 uplink throughput, which the loss itself reduces. On F1-C, loss is within the top three in 31 of 40 incidents. The nine misses are low- or medium-strength attacks in which only 7 to 23 packets were lost over the whole attack. Even when the first metric is not the ground truth, it belongs to the attacked interface in 264 of the 280 incidents.}

\subsubsection{\rev{Stability of the Causal Graphs}}
\label{subsec:eval-stability}

\rev{We also evaluate how the graphs of Tool~\#1 react to the number of averaged SAM runs, edge threshold value and noise in the telemetry.}

\rev{\textbf{Number of SAM runs.} To measure how much SAM's own randomness changes the graph, we split the 16 SAM runs of each incident into two halves of eight and average each half separately. The two resulting graphs share a median of 68\% of their edges (Jaccard index 0.68). Averaging $M=4$, $8$ or $16$ runs gives top-3 accuracy of 96.2\%, 96.6\% and 96.8\%, and graphs that share 72\% and 83\% of their edges with the 16-run graph.}

\rev{\textbf{Edge threshold value.} We also study the effect of the edge threshold value in all 280 incidents (Table~\ref{tab:stability-threshold}). A low threshold adds edges that the projection must remove (622 at a threshold of $0.3$) and F-node edges that point away from the attacked interface (23\% of them at $0.3$). A high threshold leaves more incidents with no F-node edge (21 at $0.7$). A threshold of $0.5$ balances the two.}

\begin{table}[t]
\centering
\footnotesize
\setlength{\tabcolsep}{4pt}
\caption{Effect of the edge threshold over the 280 incidents}
\label{tab:stability-threshold}
\begin{tabular}{cccccc}
\toprule
Thr. & Edges & Removed & \shortstack{No\\F edge} & \shortstack{F edges\\on target} & Overlap \\
\midrule
0.3 & 11 & 622 & 1  & 77\% & 0.75 \\
0.4 & 9  & 271 & 5  & 86\% & 0.88 \\
0.5 & 8  & 48  & 9  & 92\% & 1.00 \\
0.6 & 6  & 0   & 12 & 95\% & 0.78 \\
0.7 & 5  & 0   & 21 & 97\% & 0.62 \\
\bottomrule
\end{tabular}
\end{table}

\rev{\noindent\textbf{Noisy telemetry and sparsity weight.} On one incident of each of the 28 attack configurations, we retrained SAM after adding Gaussian noise to every metric, with a standard deviation of $0.1$ and $0.3$ times the metric's own, and with the sparsity weight $\lambda_S$ halved and doubled (Table~\ref{tab:stability-perturb}). Across the 112 retrained graphs (28 per tested case), the localisation changed in one incident: with $\lambda_S$ halved, an E2 packet-loss incident lost its top-1 hit (the ground truth moved to rank 2). Top-3 accuracy stayed at the reference value of 27 of 28 in every condition (the one miss is an F1-C packet-loss incident with 23 lost packets), and the F node kept the same children in the median incident. Under the weaker noise and both sparsity weights, the edges between metrics changed about as much as between two halves of SAM's own runs (median Jaccard index of 0.67 to 0.70, against 0.67 for these incidents); under the stronger noise of $0.3\sigma$ they changed more (0.55).}

\begin{table}[t]
\centering
\caption{Retraining under noise and a different sparsity weight (28 incidents, one per attack configuration).}
\label{tab:stability-perturb}
\begin{tabular}{lcccc}
\toprule
Condition & Top-1 & Top-3 & Edges & Overlap \\
\midrule
Reference & 26 & 27 & 9 & -- \\
Noise, $0.1\sigma$ & 26 & 27 & 8 & 0.67 \\
Noise, $0.3\sigma$ & 26 & 27 & 8 & 0.55 \\
$\lambda_S \times 0.5$ & 25 & 27 & 10 & 0.70 \\
$\lambda_S \times 2$ & 26 & 27 & 7 & 0.67 \\
\bottomrule
\end{tabular}
\end{table}

\rev{Graphs of repeats of the same configuration share a median of 50\% of their edges. We conclude that the ranking and the edges leaving $\mathcal{F}$ are stable, while individual edges between metrics vary about as much as SAM's own randomness, and more under strong noise; the analyst should read the latter as plausible propagation paths rather than as established causal links.}

\subsection{\system{}'s Tool~\#2 Evaluation}
\label{subsec:eval-query-graph-tool}

\rev{To train the soft-token encoding (Sec.~\ref{subsubsec:training}), we built the \emph{O-RAN Causal Inference Questions and Answers} (O-CIQA) dataset from the 280 per-incident DAGs of Tool~\#1 (Sec.~\ref{subsubsec:dataset}), with \rev{24 questions per graph from ten templates} (see Table~\ref{tab:ociqa}): 16 graph-level questions (the number of edges; the number and list of affected metrics; whether every metric connected to $\mathcal{F}$ is also reachable from it; whether, and with how many metrics, each interface and each O-RAN component is affected) and 8 node-level questions about $\mathcal{F}$ (the number of affected interfaces, the children of $\mathcal{F}$ per interface, and the number of children per metric type). A metric is \emph{affected} when it is reachable from $\mathcal{F}$ in the DAG, and every answer is computed from the graph. We split the O-CIQA dataset as follows: for each attack in the dataset we keep 6 repeats for the training set, 2 repeats for the validation set and 2 repeats for the test set.} \rev{We also pre-train the encoder on GraphQA~\cite{fatemi2023talk} tasks (node count, edge count, cycle check and triangle counting; 3,928 training questions) and then fine-tune it on O-CIQA.} 

\begin{table}[t!]
\centering

\small
\begin{tabular}{p{0.66\linewidth}cc}
\toprule
\textbf{Question template} & \textbf{Level} & \textbf{\#} \\
\midrule
What is the total number of edges in this graph? & graph & 1 \\
How many KPIs are affected by the anomaly, i.e., reachable from $\mathcal{F}$? & graph & 1 \\
List the KPIs affected by the anomaly in alphabetical order. & graph & 1 \\
Is every KPI connected to $\mathcal{F}$ (ignoring edge direction) also reachable from $\mathcal{F}$? & graph & 1 \\
Is the \{I\} interface affected by the anomaly? & graph & 4 \\
How many \{I\} interface KPIs are affected by the anomaly? & graph & 4 \\
Is the \{C\} component affected by the anomaly? & graph & 4 \\
How many distinct O-RAN interfaces contain at least one KPI reachable from $\mathcal{F}$? & node & 1 \\
List the \{I\} interface KPIs directly connected to $\mathcal{F}$, in alphabetical order. & node & 4 \\
How many \{M\} KPIs are directly connected to $\mathcal{F}$? & node & 3 \\
\midrule
\textbf{Total} (16 graph-level, 8 node-level) & & 24 \\
\bottomrule
\end{tabular}
\caption{\rev{The ten O-CIQA question templates and the number of questions each gives per graph (24 in total). \{I\} stands for each of the four interfaces (A1, E2, F1-C, F1-U), \{C\} for each of the four components (O-DU, O-CU, near-RT RIC, non-RT RIC) and \{M\} for each of the three metric types (latency, packet loss, retransmissions). Component questions also state which interfaces make the component affected.}}
\label{tab:ociqa}
\end{table}

\rev{\textbf{Structural queries evaluation:} To begin with, we evaluate the structural queries that the agent sends to Tool~\#2 in the verification phase. We use the O-CIQA training split to train the soft-token encoder, the validation split to stop the training and select the checkpoint (Sec.~\ref{subsubsec:training}), and the test split to report the results. A question counts as correct when the model's first answer is correct: the same number for counting questions, the same yes or no for binary questions, or the same set of metrics for listing questions; for counting questions we also report the RMSE. We compare four configurations with the same frozen LLM: the LLM given the graph as text only, without the soft token (\textit{Text only}); the GraphQA pre-trained encoder without O-CIQA training (\textit{GIN zero-shot}); and the encoder trained on O-CIQA from random weights (\textit{GIN~FS}) or from the GraphQA pre-trained weights (\textit{GIN~FT}). We evaluate each configuration with the base \textit{Llama-3.1-8B} and with its instruction-tuned variant, \textit{Llama-3.1-8B-Instruct}. As a graph-blind reference, we answer every question with the most common answer of its template in the training split.}

\begin{figure}[t!]
    \centering
    \includegraphics[width=1\linewidth]{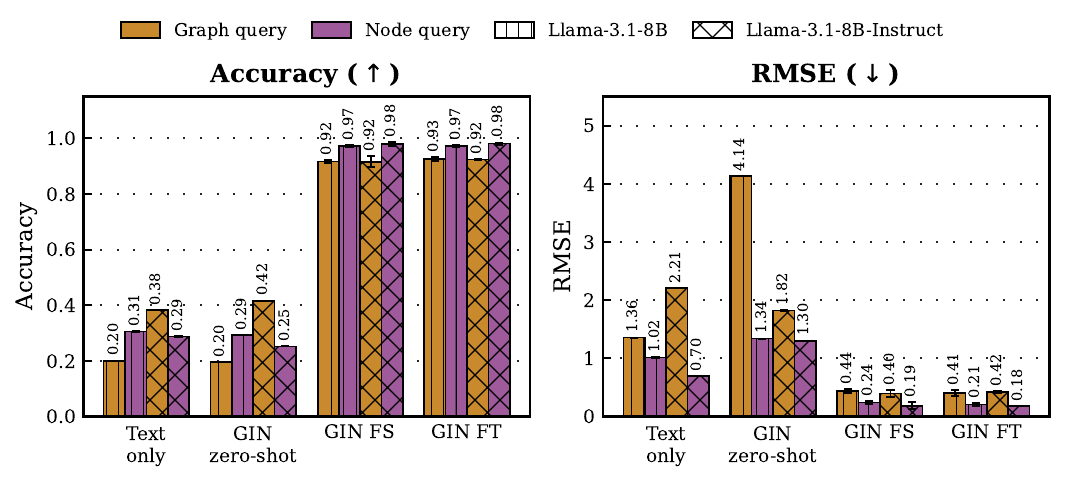}
    \caption{\rev{Graph-query accuracy (left) and RMSE of the counting questions (right) on the test set, with the frozen Llama-3.1-8B and Llama-3.1-8B-Instruct. Text only: the LLM given the edge list, without the soft token. GIN zero-shot: GraphQA pre-trained encoder without O-CIQA training. GIN FS / FT: encoder trained on O-CIQA from random or from GraphQA pre-trained weights; bars show the mean over three training seeds, error bars the standard deviation.}}
    \label{fig:graph_soft_prompting_results}
    \vspace{-11mm}
\end{figure}

\rev{Fig.~\ref{fig:graph_soft_prompting_results} shows the results. Training on O-CIQA is what makes the graph queries reliable. With Llama-3.1-8B-Instruct, \textit{GIN~FT} answers $92.4\%$ of the graph-level and $98\%$ of the node-level questions correctly over three training seeds, against $59.9\%$ and $69\%$ for the most common answer. Without O-CIQA training, the model does worse than this graph-blind reference: $38\%$ and $29\%$ for the LLM given the graph as text only, and $42\%$ and $25\%$ for the encoder pre-trained on GraphQA alone (\textit{GIN zero-shot}). The counting errors fall accordingly, from an RMSE of $2.21$ (graph level) and $0.70$ (node level) with the text-only to $0.42$ and $0.18$ with \textit{GIN~FT}. Pre-training the encoder on GraphQA brings no measurable gain: \textit{GIN~FS} reaches $92\%$ and $98\%$, within the variation across seeds. Once the encoder is trained on O-CIQA, the choice of LLM matters little: with the base Llama-3.1-8B, \textit{GIN~FT} reaches $93\%$ and $97\%$, whereas the base model given the graph as text only answers $20\%$ of the graph-level questions.}

\rev{Per question, for one trained \textit{GIN~FT} model, the questions about interfaces and components are answered almost perfectly (97.8\% each). The hardest questions are those that require all the metrics reachable from $\mathcal{F}$, i.e., following paths rather than single edges: their number is correct in 69.6\% of the test graphs and their list in 80.4\%. Whether every metric connected to $\mathcal{F}$ is also reachable from it depends on edge direction; it is answered correctly in 76.8\% of the graphs, close to the 69.6\% of the most common answer.}

\rev{\noindent\textbf{Shuffled metric identities.} The soft token depends on the graph structure only, so metric names reach the model through the prompt text or through regularities of the training incidents. To measure this, we permuted the 40 metric names of every test graph (three random permutations per graph), keeping the structure, the gates and $\mathcal{F}$, recomputed the answers and asked all questions again. With \textit{GIN~FT}, accuracy drops from $92\%$ to $74\%$ (graph level) and from $98\%$ to $88\%$ (node level), still well above the most common answer ($51\%$ and $50\%$) and the LLM given the graph as text only ($41\%$ and $17\%$). For one trained model, the number of edges is still answered correctly in $98\%$ of the cases, but counting and listing all affected metrics drop to $36.9\%$ and $44.0\%$: in real incidents the affected metrics form familiar groups, which the model partly recognizes by name instead of following the paths. Tool~\#2 is therefore used for structural questions about the incident, and the agent takes metric names from the ranked list of Tool~\#1.}

\rev{\noindent\textbf{Cost.} On one NVIDIA GH200 GPU (batch size 1, bfloat16), a query takes a median of 0.13\,s (0.09 to 1.9\,s over the 1,344 test queries), of which 3.3\,ms are the positional encoding and the GIN. The peak GPU memory is 15.1\,GB, almost all of it is the LLM. The GIN has 0.56\,M parameters. The cost is dominated by the LLM and grows with the prompt length, i.e., with the number of edges listed in the prompt.}

\subsection{\system{} Performance}
\label{subsec:eval-complete-pipeline}

\begin{figure}[t!]
    \centering
    \includegraphics[width=0.95\linewidth]{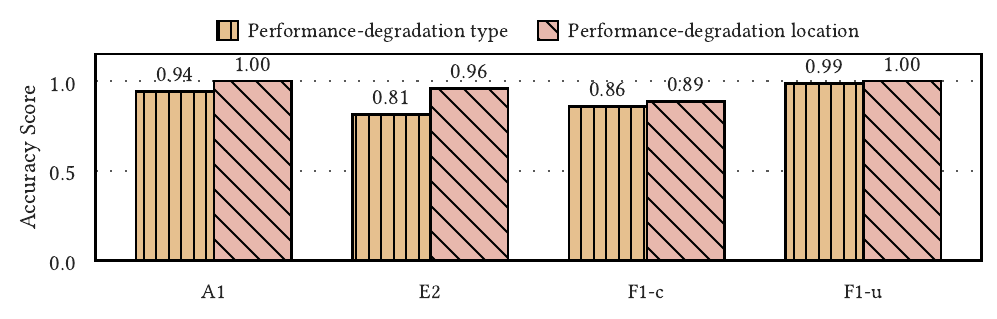}
    \caption{\rev{Classification accuracy of \system{} per attacked interface (70 incidents each: 30 delay, 40 loss).}}
    \label{fig:aggregate_clasification_accuracy}
\end{figure}

\begin{figure*}[t!]
    \centering
    \begin{subfigure}[b]{0.49\linewidth}
        \centering
        \includegraphics[width=\linewidth]{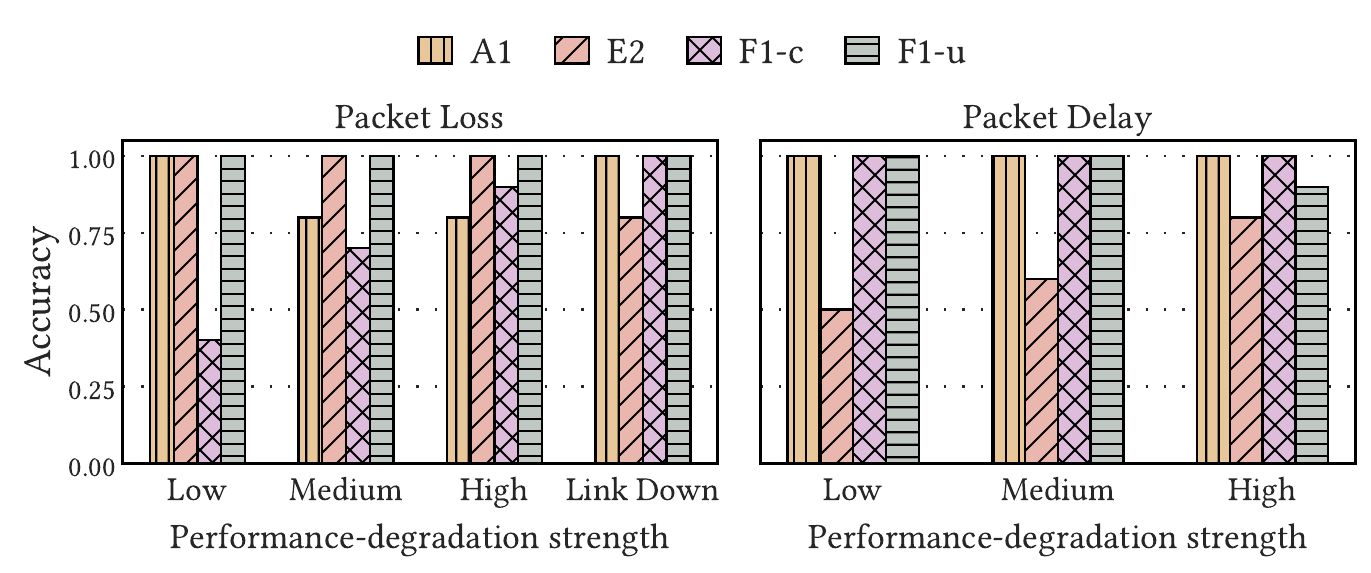}
        \caption{\rev{Degradation type (delay or loss).}}
        \label{fig:anomaly_type_final_cls}
    \end{subfigure}
    \begin{subfigure}[b]{0.48\linewidth}
        \centering
        \includegraphics[width=\linewidth]{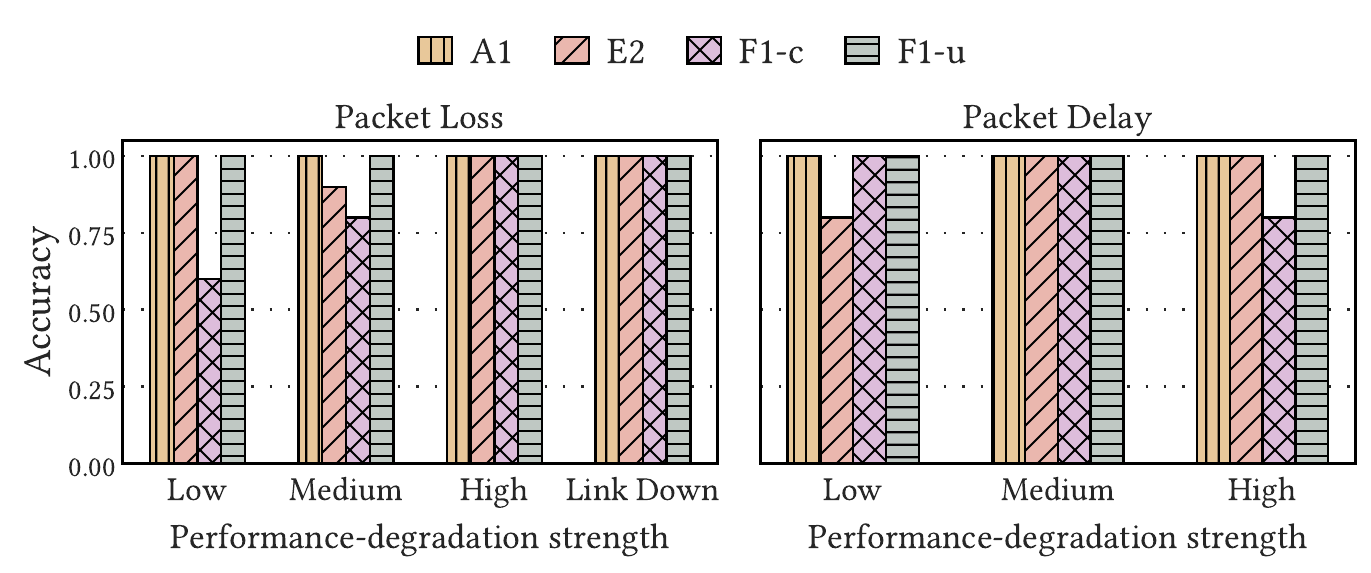}
        \caption{\rev{Degradation location (attacked interface).}}
        \label{fig:anomaly_location_final_cls}
    \end{subfigure}
    \caption{\rev{Classification accuracy of \system{} per attacked interface and attack strength. Each bar covers the 10 incidents of one attack configuration.}}
    \label{fig:perception-agent-cls-strength}

\end{figure*}

This section evaluates the complete agentic pipeline, characterizing its ability to correctly identify the type and location (interface and component) of performance-degradation incidents across the four tested O-RAN interfaces (A1, F1-C, F1-U, and E2). To systematically evaluate the text-based final reports ($\mathcal{R}$), an external LLM node processes each report to extract a structured response. The target fields for this extraction---degradation type and incident location---are strictly constrained to a predefined set of values matching the ground-truth parameters established during the experimental incident generation.

Figure~\ref{fig:aggregate_clasification_accuracy} reports the aggregate classification accuracy of \system{} across all adversarial perturbations, broken down by interface and prediction objective. \rev{Over the 280 incidents, \system{} identifies the type of the degradation in 90\% of the incidents (108 of 120 delay and 144 of 160 loss incidents) and its interface in 96\%. Every incident is counted. An incident for which the agent names no type or no interface counts as an error. Type identification is lowest on E2 (81\%), where the report names the correct metric but the structured classification is left as ``unknown'' (Sec.~\ref{subsubsec:misclassification-pipeline}), and location identification is lowest on F1-C (89\%).}

\subsubsection{\system{}'s Accuracy}
\label{subsubsec:accuracy-pipeline}

Figure~\ref{fig:perception-agent-cls-strength} disaggregates the accuracy of \system{} across two primary prediction objectives: identifying the performance-degradation type and pinpointing its location. Within each objective, the \textit{x}-axis represents the strength of the degradation event. \rev{The strengths are those of Sec.~\ref{subsubsec:dataset}: for packet loss, ``Low'' corresponds to 12--30\%, ``Medium'' to 40--60\%, ``High'' to 70--89\% loss, and ``Link Down'' to a complete interface failure; for packet delay, ``Low'' corresponds to 10--94\,ms, ``Medium'' to 111--992\,ms, and ``High'' to 1.07--4.91\,s.} Results are further grouped by the affected interface.

\rev{\noindent\textbf{A1 Interface.} Location is identified in all 70 incidents. The type is correct in all delay incidents and in 36 of 40 loss incidents. The four errors, at medium and high loss, are classified as delay.}

\rev{\noindent\textbf{E2 Interface.} Loss incidents are typed correctly in 38 of 40 cases (the two errors are link-down incidents classified as delay). Delay incidents are typed correctly in only 19 of 30 (5, 6 and 8 of 10 at low, medium and high strength): in the other 11 the report names the E2 uplink latency as the root cause, but the structured classification is left as ``unknown''. Location is correct in 67 of 70 incidents.}

\rev{\noindent\textbf{F1-C Interface.} Delay incidents are typed correctly in all 30 cases, and loss incidents in 30 of 40; the errors concentrate at low and medium strength (4 and 7 of 10 correct), where F1-C, which carries about one packet per second, loses only 7 to 23 packets during the whole attack. Location is correct in 62 of 70 incidents; the errors are on low- and medium-strength loss (6) and high-strength delay (2), and the wrong locations are F1-U (5) and A1 (3).}

\rev{\noindent\textbf{F1-U Interface.} Type and location are correct in all incidents except one high-strength delay incident typed as loss.}

\begin{figure}[t!]
    \centering
    \includegraphics[width=0.95\linewidth]{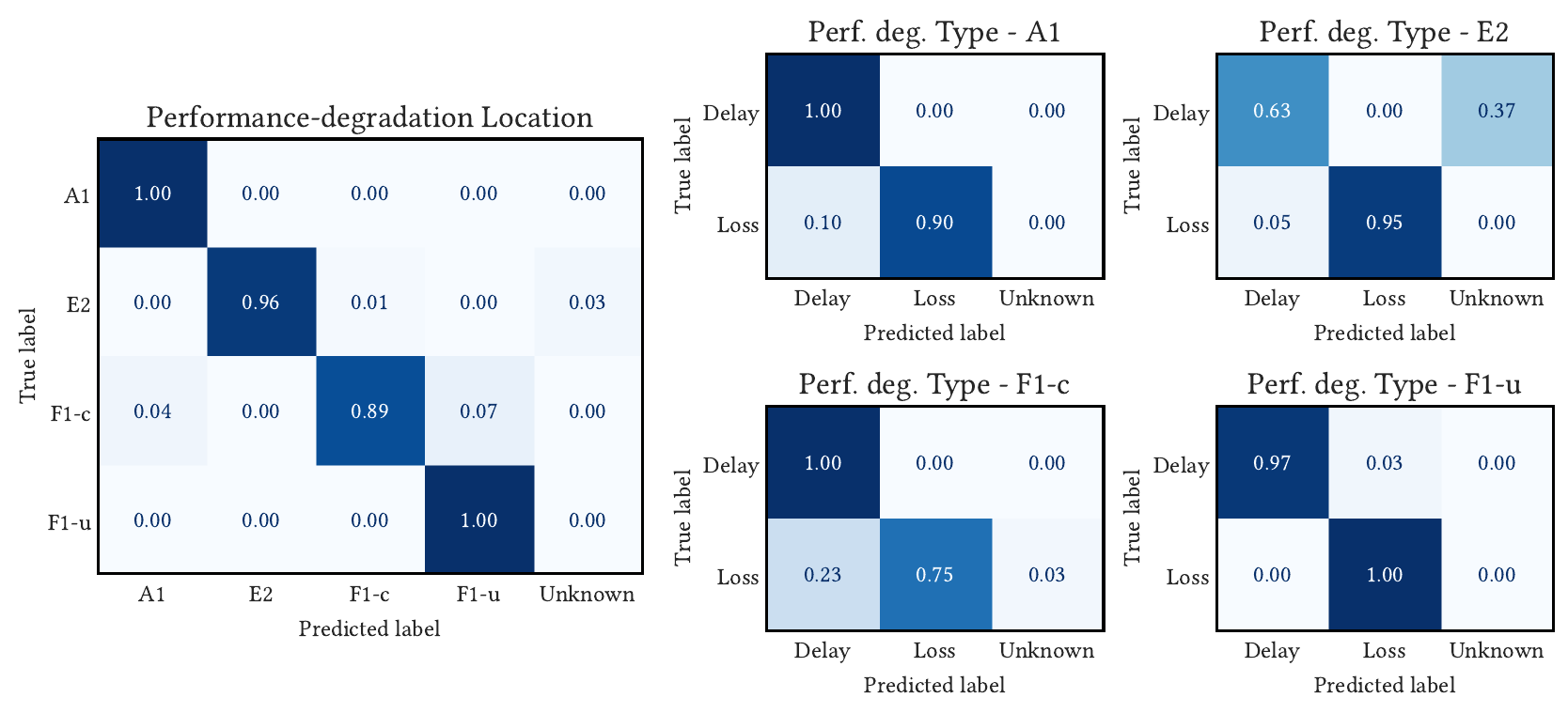}
    \caption{Confusion matrices for performance-degradation location (left) and performance-degradation type per interface (right). \rev{The column ``Unknown'' counts the incidents for which the agent named no type or location; each row is normalized over all incidents of its true label.}}
    \label{fig:confusion_matrices}
\end{figure}

\subsubsection{\system{}'s Misclassifications}
\label{subsubsec:misclassification-pipeline}

To further analyze the performance of \system{}, we decompose the misclassification structure to identify where the metric space is incomplete. Figure~\ref{fig:confusion_matrices} depicts the confusion matrices for performance-degradation location and type identification across all attacked interfaces. \rev{The location matrix (left) has diagonal values between 0.89 (F1-C) and 1.00 (A1, F1-U); F1-C incidents are assigned to F1-U (5) or A1 (3), and two E2 incidents have no location and one is assigned to F1-C.} The four performance-degradation type matrices (right) reveal \rev{two patterns}:

\rev{\noindent\textbf{F1-C packet loss with few lost packets.} 10 of the 40 F1-C loss incidents are not typed as loss (9 as delay, 1 unknown), six of them at the lowest strength. F1-C carries about one packet per second, so these attacks drop only 7 to 23 packets, and the measurements carry little evidence of loss: in five of the ten incidents the causal graph has no edge from $\mathcal{F}$, in three more $\mathcal{F}$ is connected only to throughput metrics of other interfaces, and eight of the ten are among the nine incidents in which Tool~\#1 does not rank the packet loss within the top three (Sec.~\ref{subsec:eval-causal-ranked-list}).}

\rev{\noindent\textbf{E2 delay left unclassified.} In 11 of the 30 E2 delay incidents the classification is ``unknown'', although the report names the E2 uplink latency as the root cause and the causal graph connects $\mathcal{F}$ to exactly the two E2 uplink latency metrics. The type is lost in the final step that converts the report into the structured classification, not in the diagnosis.}

\subsection{Grounding of the Reports}
\label{subsec:eval-final-report}

\rev{Similarity to a reference report would not show whether the evidence cited in a report exists. We therefore check every checkable claim of the 280 reports against the causal graph $\mathcal{G}_W$ and the ranked list $\mathcal{L}$ of the same incident, with a script and without any language model involved. The checks cover five kinds of claims: ($i$)~the metric names cited in the report, which must be among the 40 metrics; ($ii$)~the ranks and scores quoted from $\mathcal{L}$; ($iii$)~the metrics listed as directly connected to $\mathcal{F}$, which must be children of $\mathcal{F}$ in $\mathcal{G}_W$; ($iv$)~the metrics of the failure chain, which must be reachable from $\mathcal{F}$; and ($v$)~the affected interfaces and components, which must match $\mathcal{G}_W$ under the definitions of O-CIQA (Sec.~\ref{subsec:eval-query-graph-tool}).}

\rev{The claims about metrics are almost always grounded. Of the 650 metric names in the reports, 645 are among the 40 metrics, and the other five are shortened forms of existing names. All 565 ranks and scores quoted from the ranked list match it, and 360 of the 366 metrics in the failure chains are reachable from $\mathcal{F}$. Of the 616 metrics listed as directly connected to $\mathcal{F}$, 534 are children of $\mathcal{F}$; most of the others are second- or third-ranked metrics reached through another metric, which the agent lists under the report heading that the prompt reserves for metrics directly connected to $\mathcal{F}$ (Appendix~\ref{subsec:appendix-input-output-agent}).}

\rev{The claims about interfaces and components are less reliable. The affected interfaces match the graph in 262 reports and the affected components in 209. In 56 of the 62 reports that miss an affected component, Tool~\#2 had answered the agent's question about that component wrongly: the agent asked it without the sentence that defines when a component is affected, a form that Tool~\#2 answers correctly in only 42\% of the cases, against 98\% for the trained forms (Table~\ref{tab:ociqa}).}

\rev{Overall, 178 of the 280 reports (64\%) pass every check, and 226 (81\%) pass all checks on the metrics they cite, i.e., checks ($i$)--($iv$). The errors that remain come from the structural queries and from the report template rather than from invented evidence, and each of them can be detected by the analyst with the same check.}

\subsection{Execution time}
\label{subsec:eval-execution-time}
\begin{figure}[t!]
    \centering
    \includegraphics[width=0.55\columnwidth]{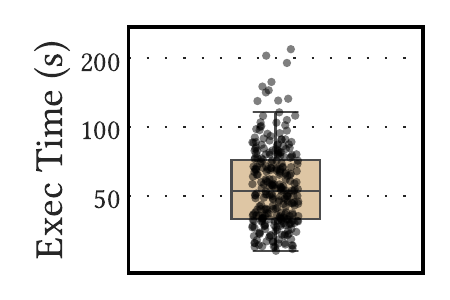}
    \caption{\rev{End-to-end execution time of the agent per incident (280 incidents; median 53\,s).}}
    \label{fig:execution_time}
\end{figure}

\rev{Figure~\ref{fig:execution_time} reports the end-to-end execution time of the agent, including its queries to Tool~\#2, over the 280 incidents. The agent takes a median of 53\,s per incident (40 to 71\,s for the middle half, at most 218\,s), most of it in its own LLM: it sends a median of 8 queries to Tool~\#2 (2 to 44), about one second in total, and its time grows with the number of tool calls it decides to make (correlation 0.72), which is highest on F1-C and F1-U. Tool~\#1 takes a median of 1.5 minutes per incident, with the 16 SAM runs trained concurrently on one NVIDIA GH200 GPU under MPS (Sec.~\ref{subsec:experimental-setup}); adding it, an incident is analysed in about two and a half minutes. Incidents are processed independently, so throughput grows with the number of model instances.}

\section{Related Work}
\label{sec:back:related}

\rev{\textbf{Anomaly and attack detection in O-RAN.} SpotLight~\cite{sun2024spotlight} detects anomalies in O-RAN metrics and uses causal discovery to identify the KPIs that caused them, DET-RAN~\cite{scalingi2024det} detects attacks in real time from cross-layer RAN data, and DAST~\cite{dast} combines a vision-language model and an LLM to detect anomalies across O-RAN interfaces on the testbed used in this work. \system{} starts from the alert of such a detector: it localizes the degradation across interfaces with a causal graph of the incident and gives the analyst the evidence to route the ticket to network maintenance or security operations.}

\rev{\textbf{Causal root-cause analysis.} RCD~\cite{ikram2022root} represents a failure as an intervention on a failure-indicator node and localizes its root cause with conditional-independence tests, and CIRCA~\cite{li2022circa} combines a graph derived from the system architecture with regression-based hypothesis testing. Both target microservices and output a ranked list of suspects. \system{} adopts RCD's failure-indicator node, adds a node for the UE population, learns one causal DAG per incident, measures its stability (Sec.~\ref{subsec:eval-stability}), and uses the graph as evidence that an agent can query.}

\rev{\textbf{LLM-based incident and network diagnosis.} RCACopilot~\cite{chen2024rcacopilot} lets an LLM predict the root-cause category of cloud incidents from diagnostic data collected by engineer-written handlers, BiAn~\cite{wang2025bian} ranks the likely failed devices of a production network from monitoring data, and Wang et al.~\cite{wang2025graphllm} guide an LLM with an alarm graph to localize the root-cause alarm of an optical network. In these systems the LLM reasons over the diagnostic data itself. In \system{}, raw telemetry never enters the prompt and the LLM never builds or changes the causal graph, so each claim of the report can be checked against a graph edge or a measurement (Sec.~\ref{subsec:eval-final-report}).}

\rev{\textbf{Graph-enhanced LLMs.} GraphToken~\cite{perozzi2024graphtoken} encodes a graph into a soft token that lets a frozen LLM answer structural questions, and G-Retriever~\cite{he2024gretriever} passes a retrieved subgraph to a frozen LLM as a soft prompt, noting that a single graph embedding cannot recall the full graph structure; we observe a related limit when metric names are permuted (Sec.~\ref{subsec:eval-query-graph-tool}). CausalGraph2LLM~\cite{sheth2025causalgraph2llm} shows that LLM answers to causal-graph queries are highly sensitive to the text encoding of the graph, which motivates a learned encoder. These works use benchmark or synthetic graphs; \system{} applies soft prompting to the causal graphs of real network incidents inside an agent.}

\section{\rev{Limitations and Future Work}}
\label{sec:future-work}
\rev{Our evaluation covers packet delay and packet loss injected on four interfaces of one testbed, with one attack at a time. It measures where the degradation enters and what type it is, not the routing decision: all incidents are injected degradations, so it does not show how \system{} behaves on benign faults that resemble attacks, such as misconfigurations or congestion. Nor does it measure whether the recommended diagnostic actions are executable and whether the reports lead analysts to route tickets correctly, which would require executing the actions on the testbed or a study with NOC analysts. \rev{The reports also carry no timing information: in our incidents the top-ranked metric starts deviating within a few seconds of the incident start, so we did not report onsets, but they may help locate cascading failures in other incidents}. Addressing these requires a substantially broader dataset with benign incidents, more attack patterns, varying network conditions and several deployments, which we leave to future work.}
\section{Conclusion}
\label{sec:conclusion}

In this paper, we presented \system{}, an agentic system that \rev{supports the triage of} O-RAN incidents. Operationally, \system{} targets the repetitive, judgment-bound triage tasks of standard performance-degradation incidents. By converting raw KPI-deviation alerts into incident reports \rev{whose evidence can be checked}, the pipeline replaces manual overhead with auditable inference, preserving the critical human-in-the-loop boundary required for network reconfiguration. The system models performance-degradation incidents as weighted causal graphs, compresses them into continuous soft tokens for a frozen LLM, and drives an autonomous loop to produce structured incident triage reports. We evaluated \system{} on a physical O-RAN deployment\rev{, over 280 performance-degradation incidents}. \rev{The causal ranking places the ground-truth metric within the top three candidates in $97\%$ of the incidents (first in $90\%$). The soft token lets the LLM answer $92\%$ of the structural queries about unseen incident graphs correctly ($38\%$ with the graph given as text only), and the agent identifies the type of the degradation in $90\%$ of the incidents and its location in $96\%$.}
The generated report, $\mathcal{R}$, functions as a verifiable diagnostic artifact. \rev{Its claims can be checked against the measurements and the edges of $\mathcal{G}_W$, enabling operators to validate or reject any reasoning step; in our evaluation, $64\%$ of the reports pass every such check and $81\%$ pass all checks on the metrics they cite (Sec.~\ref{subsec:eval-final-report}).}

\section*{ACKNOWLEDGMENT}
This work was supported in part by the ORIGAMI Project under Grant 101139270; in part by the CERCA Programme from the Generalitat de Catalunya through the ICREA programme; and in part by the funding received from Departament de Recerca I Universitats, Generalitat de Catalunya for this project

\bibliographystyle{IEEEtran}
\bibliography{references_v2}

@IEEEtranBSTCTL{BSTcontrol,
  CTLdash_repeated_names = "no",
}

@inproceedings{wang2024large,
  title={Large language models can provide accurate and interpretable incident triage},
  author={Wang, Zexin and Li, Jianhui and Ma, Minghua and Li, Ze and Kang, Yu and Zhang, Chaoyun and Bansal, Chetan and Chintalapati, Murali and Rajmohan, Saravan and Lin, Qingwei and others},
  booktitle={2024 IEEE 35th International Symposium on Software Reliability Engineering (ISSRE)},
  pages={523--534},
  year={2024},
  organization={IEEE}
}

@article{Liu_Zhang_Qian_Ma_Qin_Bansal_Lin_Rajmohan_Zhang_2024,
  author  = {Liu, Jun and Zhang, Chaoyun and Qian, Jiaxu and Ma, Minghua and Qin, Si and Bansal, Chetan and Lin, Qingwei and Rajmohan, Saravan and Zhang, Dongmei},
  title   = {Large language models can deliver accurate and interpretable time series anomaly detection},
  journal = {arXiv preprint arXiv:2405.15370},
  year    = {2024}
}

@inproceedings{Wu_Wang_Qiao_Wang_Jiang_Cui_Wang_2024,
  author    = {Wu, Duo and Wang, Xianda and Qiao, Yaqi and Wang, Zhi and Jiang, Junchen and Cui, Shuguang and Wang, Fangxin},
  title     = {{NetLLM}: Adapting large language models for networking},
  booktitle = {Proc. ACM SIGCOMM},
  pages     = {661--678},
  year      = {2024},
  doi       = {10.1145/3651890.3672268}
}

@article{zhang2023meta,
  title={Meta prompting for {AI} systems},
  author={Zhang, Yifan and Yuan, Yang and Yao, Andrew Chi-Chih},
  journal={arXiv preprint arXiv:2311.11482},
  year={2023}
}

@article{wei2022chain,
  title={Chain-of-thought prompting elicits reasoning in large language models},
  author={Wei, Jason and Wang, Xuezhi and Schuurmans, Dale and Bosma, Maarten and Xia, Fei and Chi, Ed and Le, Quoc V and Zhou, Denny and others},
  journal={Advances in Neural Information Processing Systems},
  volume={35},
  pages={24824--24837},
  year={2022}
}

@misc{fujitsu2022vran,
  author       = {{Fujitsu}},
  title        = {{Fujitsu, KDDI} Successfully Turns on the World's First {5G} Standalone Open {RAN} Site Powered by {vRAN} in {Japan}},
  howpublished = {News Release},
  year         = {2022},
  url          = {https://www.fujitsu.com/global/about/resources/news/press-releases/2022/0218-01.html}
}

@misc{nttdocomo2023openran,
  author       = {{NTT DOCOMO}},
  title        = {{NTT DOCOMO} Offers Open {RAN} Product Set to Other Telcos},
  howpublished = {Fierce Network, News Release},
  year         = {2023},
  url          = {https://www.fierce-network.com/tech/ntt-docomo-offers-open-ran-product-set-other-telcos}
}

@misc{att2023openran,
  author       = {{AT\&T}},
  title        = {{AT\&T} to Accelerate Open and Interoperable Radio Access Networks ({RAN}) in the {United States} Through New Collaboration with {Ericsson}},
  howpublished = {News Release},
  year         = {2023},
  url          = {https://about.att.com/story/2023/commercial-scale-open-radio-access-network.html}
}

@misc{gsma2021openranmou,
  author       = {{GSMA}},
  title        = {Major {European} Operators Sign Open {RAN} {MoU}},
  howpublished = {News Release},
  year         = {2021},
  url          = {https://www.gsma.com/futurenetworks/digest/major-european-operators-sign-open-ran-mou/}
}

@misc{lightreading2024orange,
  author       = {{Light Reading}},
  title        = {Amarisoft Powered {Orange} {5G} at {Olympics} in Best Open {RAN} Example Yet},
  howpublished = {Light Reading},
  year         = {2024},
  url          = {https://www.lightreading.com/open-ran/amarisoft-powered-orange-5g-at-olympics-in-best-open-ran-example-yet}
}

@article{baguer2024attacking,
  author  = {P. Baguer and G. M. Yilma and E. Municio and G. Garcia-Aviles and A. Garcia-Saavedra and M. Liebsch and X. Costa-P\'{e}rez},
  title   = {Attacking {O-RAN} interfaces: Threat modeling, analysis and practical experimentation},
  journal = {IEEE Open J. Commun. Soc.},
  year    = {2024}
}

@techreport{oran-wg11-threat,
  author       = {{O-RAN Alliance}},
  title        = {Security Threat Modeling and Risk Assessment 8.0},
  institution  = {{O-RAN Alliance}},
  type         = {Technical Report},
  number       = {O-RAN.WG11.TR.Threat-Modeling-R005-v08.00},
  year         = {2026},
  note         = {{Work Group 11} (Security). Available at \url{https://www.o-ran.org/specifications}},
}

@techreport{oran-wg5-cp,
  author       = {{O-RAN Alliance}},
  title        = {{NR} {C}-plane profile 17.0},
  institution  = {{O-RAN Alliance}},
  type         = {Technical Specification},
  number       = {O-RAN.WG5.TS.C.1-R005-v17.00},
  year         = {2026},
  note         = {{Work Group 5} (Open F1/W1/E1/X2/Xn Interface). Available at \url{https://www.o-ran.org/specifications}},
}

@techreport{oran-wg3-e2gap,
  author       = {{O-RAN Alliance}},
  title        = {{O-RAN} {E2} General Aspects and Principles ({E2GAP}) 8.0},
  institution  = {{O-RAN Alliance}},
  type         = {Technical Specification},
  number       = {O-RAN.WG3.TS.E2GAP-R004-v08.00},
  year         = {2025},
  note         = {{Work Group 3} (Near-Real-Time RIC and E2 Interface). Available at \url{https://www.o-ran.org/specifications}},
}

@techreport{oran-wg2-a1gap,
  author       = {{O-RAN Alliance}},
  title        = {{A1} Interface: General Aspects and Principles ({A1GAP}) 5.03},
  institution  = {{O-RAN Alliance}},
  type         = {Technical Specification},
  number       = {O-RAN.WG2.TS.A1GAP-R005-v05.03},
  year         = {2026},
  note         = {{Work Group 2} (Non-Real-Time RIC and A1 Interface). Available at \url{https://www.o-ran.org/specifications}},
}

@inproceedings{scalingi2024det,
  title={{DET-RAN}: Data-driven cross-layer real-time attack detection in {5G} {Open RANs}},
  author={Scalingi, Alessio and D'Oro, Salvatore and Restuccia, Francesco and Melodia, Tommaso and Giustiniano, Domenico},
  booktitle={IEEE INFOCOM 2024-IEEE Conference on Computer Communications},
  pages={41--50},
  year={2024},
  organization={IEEE}
}

@inproceedings{hussain20195greasoner,
  title={{5GReasoner}: A property-directed security and privacy analysis framework for {5G} cellular network protocol},
  author={Hussain, Syed Rafiul and Echeverria, Mitziu and Karim, Imtiaz and Chowdhury, Omar and Bertino, Elisa},
  booktitle={Proceedings of the 2019 ACM SIGSAC Conference on Computer and Communications Security},
  pages={669--684},
  year={2019}
}

@inproceedings{xing2023enabling,
  title={Enabling resilience in virtualized {RANs} with {Atlas}},
  author={Xing, Jiarong and Gong, Junzhi and Foukas, Xenofon and Kalia, Anuj and Kim, Daehyeok and Kotaru, Manikanta},
  booktitle={Proc. ACM MobiCom},
  pages={1--15},
  year={2023}
}

@inproceedings{lin20255g,
  title={{5G-Muffler}: Covert {DoS} attacks over open fronthaul interface of {O-RAN} {5G} network},
  author={Lin, Wei and Li, Zongxiao and Chen, Binbin and Liu, Jianwei and Cheng, Ray-Guang and Zhang, Fan},
  booktitle={IEEE INFOCOM 2025-IEEE Conference on Computer Communications},
  pages={1--10},
  year={2025},
  organization={IEEE}
}

@inproceedings{wen20245g,
  title={{5G-Spector}: An {O-RAN} compliant layer-3 cellular attack detection service},
  author={Wen, Haohuang and Porras, Phillip A and Yegneswaran, Vinod and Gehani, Ashish and Lin, Zhiqiang},
  booktitle={Proc. NDSS},
  year={2024}
}

@inproceedings{xing2024criticality,
  title={On the criticality of integrity protection in {5G} fronthaul networks},
  author={Xing, Jiarong and Yoo, Sophia and Foukas, Xenofon and Kim, Daehyeok and Reiter, Michael K},
  booktitle={33rd USENIX Security Symposium (USENIX Security 24)},
  pages={4463--4479},
  year={2024}
}

@article{sundqvist2023robust,
  title={Robust procedural learning for anomaly detection and observability in {5G} {RAN}},
  author={Sundqvist, Tobias and Bhuyan, Monowar and Elmroth, Erik},
  journal={IEEE Transactions on Network and Service Management},
  volume={21},
  number={2},
  pages={1432--1445},
  year={2024},
  publisher={IEEE}
}

@inproceedings{groen2023cost,
  title={The cost of securing {O-RAN}},
  author={Groen, Joshua and Kim, Brian and Chowdhury, Kaushik},
  booktitle={ICC 2023-IEEE International Conference on Communications},
  pages={5444--5449},
  year={2023},
  organization={IEEE}
}

@article{groen2024securing,
  title={Securing {O-RAN} open interfaces},
  author={Groen, Joshua and D'Oro, Salvatore and Demir, Utku and Bonati, Leonardo and Villa, Davide and Polese, Michele and Melodia, Tommaso and Chowdhury, Kaushik},
  journal={IEEE Transactions on Mobile Computing},
  volume={23},
  number={12},
  pages={11265--11277},
  year={2024},
  publisher={IEEE}
}

@article{kalainathan2022structural,
  title={Structural agnostic modeling: Adversarial learning of causal graphs},
  author={Kalainathan, Diviyan and Goudet, Olivier and Guyon, Isabelle and Lopez-Paz, David and Sebag, Mich{\`e}le},
  journal={Journal of Machine Learning Research},
  volume={23},
  number={219},
  pages={1--62},
  year={2022}
}

@inproceedings{thimmaraju2024security,
  title={Security testing the {O-RAN} near-real time {RIC} \& {A1} interface},
  author={Thimmaraju, Kashyap and Shaik, Altaf and Fl{\"u}ck, Sunniva and Mora, Pere Joan Fullana and Werling, Christian and Seifert, Jean-Pierre},
  booktitle={Proceedings of the 17th ACM Conference on Security and Privacy in Wireless and Mobile Networks},
  pages={277--287},
  year={2024}
}

@inproceedings{sun2024spotlight,
  title={{SpotLight}: Accurate, explainable and efficient anomaly detection for {Open RAN}},
  author={Sun, Chuanhao and Pawar, Ujjwal and Khoja, Molham and Foukas, Xenofon and Marina, Mahesh K and Radunovic, Bozidar},
  booktitle={Proc. ACM MobiCom},
  pages={923--937},
  year={2024}
}

@inproceedings{chawla2020interpretable,
  title={Interpretable unsupervised anomaly detection for {RAN} cell trace analysis},
  author={Chawla, Ashima and Jacob, Paul and Feghhi, Saman and Rughwani, Devashish and van der Meer, Sven and Fallon, Sheila},
  booktitle={2020 16th International Conference on Network and Service Management (CNSM)},
  pages={1--5},
  year={2020},
  organization={IEEE}
}

@inproceedings{iyer2017automating,
  title={Automating diagnosis of cellular radio access network problems},
  author={Iyer, Anand Padmanabha and Li, Li Erran and Stoica, Ion},
  booktitle={Proc. ACM MobiCom},
  pages={79--87},
  year={2017}
}

@inproceedings{garcia2021nuberu,
  title={Nuberu: Reliable {RAN} virtualization in shared platforms},
  author={Garcia-Aviles, Gines and Garcia-Saavedra, Andres and Gramaglia, Marco and Costa-Perez, Xavier and Serrano, Pablo and Banchs, Albert},
  booktitle={Proc. ACM MobiCom},
  pages={749--761},
  year={2021}
}

@techreport{ORAN_WG4_Fronthaul_CUS,
  author       = {{O-RAN Alliance}},
  title        = {Control, User and Synchronization Plane Specification 20.0},
  institution  = {{O-RAN Alliance}},
  type         = {Technical Specification},
  number       = {O-RAN.WG4.TS.CUS.0-R005-v20.00},
  year         = {2026},
  note         = {{Work Group 4} (Open Fronthaul Interfaces). Available at \url{https://www.o-ran.org/specifications}},
}

@article{liu2024lost,
  author  = {N. F. Liu and K. Lin and J. Hewitt and A. Paranjape and M. Bevilacqua and F. Petroni and P. Liang},
  title   = {Lost in the middle: How language models use long contexts},
  journal = {Trans. ACL},
  volume  = {12},
  pages   = {157--173},
  year    = {2024}
}

@article{fatemi2023talk,
  author  = {B. Fatemi and J. Halcrow and B. Perozzi},
  title   = {Talk like a graph: Encoding graphs for large language models},
  journal = {arXiv preprint arXiv:2310.04560},
  year    = {2023}
}

@inproceedings{sheth2025causalgraph2llm,
  author    = {I. Sheth and B. Fatemi and M. Fritz},
  title     = {{CausalGraph2LLM}: Evaluating {LLMs} for causal queries},
  booktitle = {Findings of ACL: NAACL},
  pages     = {2076--2098},
  year      = {2025}
}

@article{perozzi2024graphtoken,
  author  = {B. Perozzi and B. Fatemi and D. Zelle and A. Tsitsulin and M. Kazemi and R. Al-Rfou and J. Halcrow},
  title   = {Let your graph do the talking: Encoding structured data for {LLMs}},
  journal = {arXiv preprint arXiv:2402.05862},
  year    = {2024}
}

@inproceedings{li2022circa,
  author    = {M. Li and Z. Li and K. Yin and X. Nie and W. Zhang and K. Sui and D. Pei},
  title     = {Causal inference-based root cause analysis for online service systems with intervention recognition},
  booktitle = {Proc. ACM KDD},
  pages     = {3230--3240},
  year      = {2022}
}

@article{dwivedi2023benchmarking,
  author  = {V. P. Dwivedi and C. K. Joshi and A. T. Luu and T. Laurent and Y. Bengio and X. Bresson},
  title   = {Benchmarking graph neural networks},
  journal = {J. Mach. Learn. Res.},
  volume  = {24},
  number  = {43},
  pages   = {1--48},
  year    = {2023}
}

@article{grattafiori2024llama3,
  author  = {A. Grattafiori and others},
  title   = {The {Llama} 3 herd of models},
  journal = {arXiv preprint arXiv:2407.21783},
  year    = {2024}
}

@inproceedings{yao2023react,
  author    = {S. Yao and J. Zhao and D. Yu and N. Du and I. Shafran and K. Narasimhan and Y. Cao},
  title     = {{ReAct}: Synergizing reasoning and acting in language models},
  booktitle = {Proc. ICLR},
  year      = {2023}
}

@inproceedings{xu2019powerful,
  author    = {K. Xu and W. Hu and J. Leskovec and S. Jegelka},
  title     = {How powerful are graph neural networks?},
  booktitle = {Proc. ICLR},
  year      = {2019}
}

@misc{mitre-fight,
  author = {{MITRE Corporation}},
  title = {{FiGHT: 5G Hierarchy of Global Adversary Threats}},
  year = {2026},
  url = {https://fight.mitre.org/},
  note = {{Accessed}: Apr. 18, 2026}
}

@inproceedings{padmanabha2018mitigating,
  title={Mitigating the latency-accuracy trade-off in mobile data analytics systems},
  author={Iyer, Anand Padmanabha and Li, Li Erran and Chowdhury, Mosharaf and Stoica, Ion},
  booktitle={Proc. ACM MobiCom},
  pages={513--528},
  year={2018}
}

@inproceedings{kotaru2023adapting,
  title={Adapting foundation models for operator data analytics},
  author={Kotaru, Manikanta},
  booktitle={Proceedings of the 22nd ACM Workshop on Hot Topics in Networks},
  pages={172--179},
  year={2023}
}

@article{ikram2022root,
  title={Root cause analysis of failures in microservices through causal discovery},
  author={Ikram, Azam and Chakraborty, Sarthak and Mitra, Subrata and Saini, Shiv and Bagchi, Saurabh and Kocaoglu, Murat},
  journal={Advances in Neural Information Processing Systems},
  volume={35},
  pages={31158--31170},
  year={2022}
}

@book{spirtes2000causation,
  title={Causation, prediction, and search},
  author={Spirtes, Peter and Glymour, Clark N and Scheines, Richard},
  year={2000},
  edition={2nd},
  publisher={MIT Press}
}

@article{goodfellow2014generative,
  title={Generative adversarial nets},
  author={Goodfellow, Ian J and Pouget-Abadie, Jean and Mirza, Mehdi and Xu, Bing and Warde-Farley, David and Ozair, Sherjil and Courville, Aaron and Bengio, Yoshua},
  journal={Advances in Neural Information Processing Systems},
  volume={27},
  year={2014}
}

@book{hastie2015statistical,
  title     = {Statistical Learning with Sparsity: The Lasso and Generalizations},
  author    = {Hastie, Trevor and Tibshirani, Robert and Wainwright, Martin},
  publisher = {CRC Press},
  address   = {Boca Raton, FL, USA},
  year      = {2015}
}

@techreport{nist-sp-800-61,
  author      = {Paul Cichonski and Tom Millar and Tim Grance and Karen Scarfone},
  title       = {{Computer Security Incident Handling Guide}},
  institution = {National Institute of Standards and Technology},
  year        = {2012},
  type        = {NIST Special Publication (SP)},
  number      = {800-61 Rev. 2},
  address     = {Gaithersburg, MD},
  doi         = {10.6028/NIST.SP.800-61r2}
}

@inproceedings{pei2025flow,
  title={Flow-of-Action: {SOP} enhanced {LLM}-based multi-agent system for root cause analysis},
  author={Pei, Changhua and Wang, Zexin and Liu, Fengrui and Li, Zeyan and Liu, Yang and He, Xiao and Kang, Rong and Zhang, Tieying and Chen, Jianjun and Li, Jianhui and others},
  booktitle={Companion Proceedings of the ACM on Web Conference 2025},
  pages={422--431},
  year={2025}
}

@online{langgraph,
  author       = {{LangChain}},
  title        = {{LangGraph}},
  year         = {2024},
  url          = {https://www.langchain.com/langgraph},
  urldate      = {2026-04-29}
}

@InProceedings{Falkenberg2019a,
	Author = {Robert Falkenberg and Christian Wietfeld},
	Title = {{FALCON}: An Accurate Real-time Monitor for Client-based Mobile Network Data Analytics},
	Booktitle = {2019 IEEE Global Communications Conference (GLOBECOM)},
	Year = {2019},
	Address = {Waikoloa, Hawaii, USA},
	Month = dec,
	Publisher = {IEEE},
	Doi = {10.1109/GLOBECOM38437.2019.9014096},
	Eprint = {1907.10110},
	Eprinttype = {arxiv},
	Url = {https://arxiv.org/abs/1907.10110}
}

@online{oran_sc,
  author       = {{O-RAN Software Community}},
  title        = {{O-RAN Software Community}},
  year         = {2026},
  url          = {https://o-ran-sc.org/},
  urldate      = {2026-04-29}
}

@online{srsran,
  author       = {{Software Radio Systems}},
  title        = {{srsRAN Project}},
  year         = {2026},
  url          = {https://www.srsran.com/},
  urldate      = {2026-04-29}
}

@online{open5gs,
  author       = {{Open5GS}},
  title        = {{Open5GS}},
  year         = {2026},
  url          = {https://open5gs.org/},
  urldate      = {2026-04-29}
}

@online{gemma4,
  author       = {{Google DeepMind}},
  title        = {Gemma 4},
  year         = {2026},
  url          = {https://deepmind.google/models/gemma/gemma-4/},
  urldate      = {2026-04-30}
}

@article{team2024gemma,
  title={Gemma: Open models based on {Gemini} research and technology},
  author={{Gemma Team} and Mesnard, Thomas and Hardin, Cassidy and Dadashi, Robert and Bhupatiraju, Surya and Pathak, Shreya and Sifre, Laurent and Rivi{\`e}re, Morgane and Kale, Mihir Sanjay and Love, Juliette and others},
  journal={arXiv preprint arXiv:2403.08295},
  year={2024}
}

@article{dast,
  author  = {Spinelli, Francesco and Municio, Esteban and Baguer, Pau and Garcia-Aviles, Gines and Costa-Perez, Xavier},
  title   = {{DAST}: A {VLM}-{LLM} framework for cross-interface anomaly detection in {O-RAN}},
  journal = {arXiv preprint arXiv:2606.06261},
  year    = {2026}
}

@online{mgen,
  author       = {{U.S. Naval Research Laboratory}},
  title        = {{MGEN}: {Multi-Generator} traffic generation tool},
  year         = {2026},
  url          = {https://github.com/USNavalResearchLaboratory/mgen},
  urldate      = {2026-11-19}
}

@inproceedings{chen2024rcacopilot,
  author    = {Chen, Yinfang and Xie, Huaibing and Ma, Minghua and Kang, Yu and Gao, Xin and Shi, Liu and Cao, Yunjie and Gao, Xuedong and Fan, Hao and Wen, Ming and Zeng, Jun and Ghosh, Supriyo and Zhang, Xuchao and Zhang, Chaoyun and Lin, Qingwei and Rajmohan, Saravan and Zhang, Dongmei and Xu, Tianyin},
  title     = {Automatic root cause analysis via large language models for cloud incidents},
  booktitle = {Proc. European Conference on Computer Systems (EuroSys)},
  year      = {2024},
  doi       = {10.1145/3627703.3629553}
}

@inproceedings{wang2025bian,
  author    = {Wang, Chenxu and Zhang, Xumiao and Lu, Runwei and Lin, Xianshang and Zeng, Xuan and Zhang, Xinlei and An, Zhe and Wu, Gongwei and Gao, Jiaqi and Tian, Chen and Chen, Guihai and Liu, Guyue and Liao, Yuhong and Lin, Tao and Cai, Dennis and Zhai, Ennan},
  title     = {Towards {LLM}-based failure localization in production-scale networks},
  booktitle = {Proc. ACM SIGCOMM},
  pages     = {496--511},
  year      = {2025},
  doi       = {10.1145/3718958.3750505}
}

@article{wang2025graphllm,
  author  = {Wang, Yidi and Pang, Yue and Liu, Yanli and Zhang, Yao and Zhang, Lifang and Zhang, Min and Wang, Danshi},
  title   = {Graph structure-enhanced large language model for optical network fault diagnosis: An explainable alarm root cause localization approach},
  journal = {IEEE Internet of Things Journal},
  volume  = {12},
  pages   = {31493--31510},
  year    = {2025}
}

@inproceedings{he2024gretriever,
  author    = {He, Xiaoxin and Tian, Yijun and Sun, Yifei and Chawla, Nitesh V. and Laurent, Thomas and LeCun, Yann and Bresson, Xavier and Hooi, Bryan},
  title     = {{G-Retriever}: Retrieval-augmented generation for textual graph understanding and question answering},
  booktitle = {Advances in Neural Information Processing Systems (NeurIPS)},
  volume    = {37},
  year      = {2024}
}

\appendices
\section{Illustrative Pipeline Artifacts}
\label{app:system-prompt}

\rev{This appendix follows one incident through the pipeline: \texttt{pkt\_delay\_low-e2\_upstream-itr\_9}, a low-strength delay attack on the E2 uplink (Sec.~\ref{sec:eval}). The incident belongs to the test split of O-CIQA, so the graph encoder never saw its graph during training. All artifacts below are the outputs of the revised pipeline for this incident, reproduced without editing.}

\subsection{Causal Graph and Root-Cause Ranking}
\label{subsec:appendix-causal-graph-ranked-list}

\begin{figure}[H]
    \centering
    \includegraphics[width=1\linewidth]{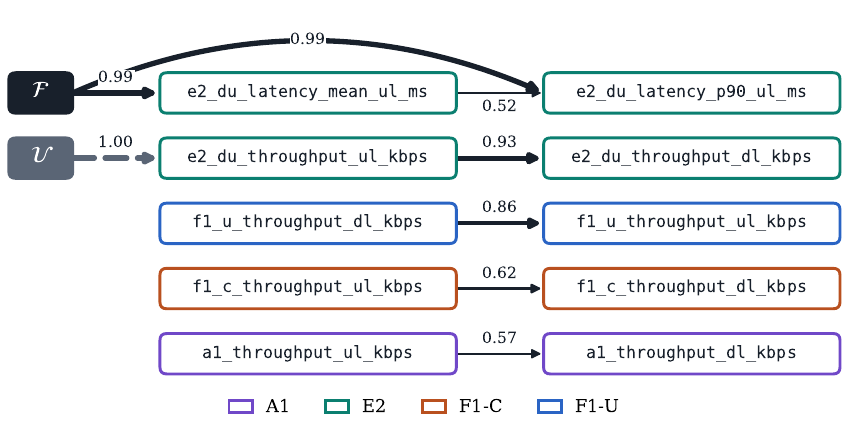}
    \caption{\rev{Causal graph $\mathcal{G}_W$ of incident \texttt{pkt\_delay\_low-e2\_upstream-itr\_9} (low-strength delay on the E2 uplink). \textbf{Box color}: interface of the metric. \textbf{Edge labels}: gate $\bar{A}_{ij}$; thicker edges have larger gates. \textbf{Dashed:} edge from $\mathcal{U}$. The 30 metrics without edges are not drawn.}}
    \label{fig:causal-graph-e2-delay-low}
\end{figure}

\rev{Fig.~\ref{fig:causal-graph-e2-delay-low} shows the causal graph $\mathcal{G}_W$ built by Tool~\#1 for this incident. It has $8$ edges. The \fnode{} has two children, the mean and the 90th-percentile uplink latency of E2, and the \uesnode{} has one, the E2 uplink throughput. The remaining edges link the uplink and downlink throughputs of each interface. The $30$ metrics that stay constant over the contextual window $C_W$ (all packet-loss and retransmission metrics, and every latency metric except the two E2 uplink ones) have no edge and are not drawn. Table~\ref{tab:anomaly-scores-example} lists the first ten metrics of the ranked list $\mathcal{L}$ (Sec.~\ref{subsubsec:ranking}). The \texttt{ranked\_list\_causes} tool returns all $40$ metrics to the agent, each with its distance $d_{\mathcal{F}}$ (reported to the agent as its score) and its rank. The two children of $\mathcal{F}$ come first, with $d_{\mathcal{F}} = 1 - \bar{A}_{\mathcal{F}v}$; every other metric is far from $\mathcal{F}$ ($d_{\mathcal{F}} > 0.8$), because its distance is computed through weak gates of $\bar{A}$ that do not reach the edge threshold.}

\begin{table}[H]
\centering
\caption{\rev{First ten metrics of the ranked list $\mathcal{L}$ for the incident of Fig.~\ref{fig:causal-graph-e2-delay-low}. Gate: $\bar{A}_{\mathcal{F}v}$ of the edge from $\mathcal{F}$, for the edges in $\mathcal{G}_W$.}}
\label{tab:anomaly-scores-example}
\footnotesize
\begin{tabular}{rlcc}
\toprule
\textbf{Rank} & \textbf{Metric} & \textbf{$d_{\mathcal{F}}$} & \textbf{Gate} \\
\midrule
1  & \texttt{e2\_du\_latency\_mean\_ul\_ms}  & 0.010 & 0.990 \\
2  & \texttt{e2\_du\_latency\_p90\_ul\_ms}   & 0.011 & 0.989 \\
3  & \texttt{f1\_u\_throughput\_dl\_kbps}    & 0.806 & -- \\
4  & \texttt{f1\_c\_throughput\_ul\_kbps}    & 0.832 & -- \\
5  & \texttt{a1\_throughput\_dl\_kbps}       & 0.859 & -- \\
6  & \texttt{e2\_du\_throughput\_dl\_kbps}   & 0.902 & -- \\
7  & \texttt{a1\_throughput\_ul\_kbps}       & 0.906 & -- \\
8  & \texttt{f1\_u\_throughput\_ul\_kbps}    & 0.929 & -- \\
9  & \texttt{f1\_c\_throughput\_dl\_kbps}    & 0.954 & -- \\
10 & \texttt{e2\_du\_throughput\_ul\_kbps}   & 0.972 & -- \\
\bottomrule
\end{tabular}
\end{table}

\subsection{Causal Reasoning: Query and Response}
\label{subsec:appendix-causal-reasoning}

\rev{During the \emph{Verify} phase of this incident, the agent sent the following five questions to the \texttt{query\_graph\_tool} (Tool~\#2). Each answer is reproduced as returned by the tool; all five agree with the answer computed from the graph of Fig.~\ref{fig:causal-graph-e2-delay-low}.}

\begin{itemize}
    \item \rev{Question: How many distinct O-RAN interfaces (F1-u, F1-c, E2, A1) contain at least one KPI that is reachable from the \fnode{}?}
    \item \rev{Answer: 1.}\newline

    \item \rev{Question: Is the E2 interface affected by the anomaly?}
    \item \rev{Answer: Yes.}\newline

    \item \rev{Question: List the E2 interface KPI nodes that are directly connected to the \fnode{}, in alphabetical order.}
    \item \rev{Answer: \texttt{e2\_du\_latency\_mean\_ul\_ms}, \texttt{e2\_du\_latency\_p90\_ul\_ms}.}\newline

    \item \rev{Question: Is the O-DU component affected by the anomaly?}
    \item \rev{Answer: Yes.}\newline

    \item \rev{Question: Is the near-RT RIC component affected by the anomaly?}
    \item \rev{Answer: Yes.}
\end{itemize}

\rev{Table~\ref{tab:appendix-ociqa-example} lists all $24$ O-CIQA questions for the same graph, the answer computed from the graph by the code that builds O-CIQA, and the answer of the fine-tuned encoder (GIN~FT) used by the agent. The prompt of every question also contains the node and directed edge list of Fig.~\ref{fig:causal-graph-e2-delay-low} as text (Sec.~\ref{subsec:soft-prompt}). GIN~FT answers all $24$ questions correctly.}

\begin{table*}[t!]
\centering
\caption{\rev{The $24$ O-CIQA questions for the graph of Fig.~\ref{fig:causal-graph-e2-delay-low}. Level: readout used (graph: mean of all nodes; node: \fnode{}). Expected: answer computed from $\mathcal{G}_W$. GIN~FT: answer of the fine-tuned encoder with Llama-3.1-8B-Instruct.}}
\label{tab:appendix-ociqa-example}
\footnotesize
\begin{tabular}{r l p{0.50\textwidth} p{0.17\textwidth} p{0.17\textwidth}}
\toprule
\textbf{\#} & \textbf{Level} & \textbf{Question} & \textbf{Expected} & \textbf{GIN~FT} \\
\midrule
1 & graph & What is the total number of edges in this graph? & 8. & 8. \\
2 & graph & How many KPIs are affected by the anomaly, i.e. how many KPI nodes are reachable from the \fnode{}? & 2. & 2. \\
3 & graph & List the KPIs affected by the anomaly in alphabetical order, i.e. list the KPI nodes that are reachable from the \fnode{} in alphabetical order. & \texttt{e2\_\allowbreak{}du\_\allowbreak{}latency\_\allowbreak{}mean\_\allowbreak{}ul\_\allowbreak{}ms}, \texttt{e2\_\allowbreak{}du\_\allowbreak{}latency\_\allowbreak{}p90\_\allowbreak{}ul\_\allowbreak{}ms}. & \texttt{e2\_\allowbreak{}du\_\allowbreak{}latency\_\allowbreak{}mean\_\allowbreak{}ul\_\allowbreak{}ms}, \texttt{e2\_\allowbreak{}du\_\allowbreak{}latency\_\allowbreak{}p90\_\allowbreak{}ul\_\allowbreak{}ms}. \\
4 & graph & Is every KPI node connected to the \fnode{} (ignoring edge direction) also reachable from the \fnode{}? & Yes. & Yes. \\
5 & node & How many distinct O-RAN interfaces (F1-u, F1-c, E2, A1) contain at least one KPI that is reachable from the \fnode{}? & 1. & 1. \\
6 & graph & Is the F1-c interface affected by the anomaly? & No. & No. \\
7 & graph & Is the F1-u interface affected by the anomaly? & No. & No. \\
8 & graph & Is the A1 interface affected by the anomaly? & No. & No. \\
9 & graph & Is the E2 interface affected by the anomaly? & Yes. & Yes. \\
10 & graph & How many F1-c interface KPIs are affected by the anomaly? & 0. & 0. \\
11 & graph & How many F1-u interface KPIs are affected by the anomaly? & 0. & 0. \\
12 & graph & How many A1 interface KPIs are affected by the anomaly? & 0. & 0. \\
13 & graph & How many E2 interface KPIs are affected by the anomaly? & 2. & 2. \\
14 & graph & Is the O-DU component affected by the anomaly? The O-DU is affected if the F1-c, F1-u or E2 interfaces are affected. & Yes. & Yes. \\
15 & graph & Is the O-CU component affected by the anomaly? The O-CU is affected if the F1-c or F1-u interfaces are affected. & No. & No. \\
16 & graph & Is the near-RT RIC component affected by the anomaly? The near-RT RIC is affected if the E2 or A1 interfaces are affected. & Yes. & Yes. \\
17 & graph & Is the non-RT RIC component affected by the anomaly? The non-RT RIC is affected if the A1 interface is affected. & No. & No. \\
18 & node & List the F1-c interface KPI nodes that are directly connected to the \fnode{}, in alphabetical order. & None. & None. \\
19 & node & List the F1-u interface KPI nodes that are directly connected to the \fnode{}, in alphabetical order. & None. & None. \\
20 & node & List the A1 interface KPI nodes that are directly connected to the \fnode{}, in alphabetical order. & None. & None. \\
21 & node & List the E2 interface KPI nodes that are directly connected to the \fnode{}, in alphabetical order. & \texttt{e2\_\allowbreak{}du\_\allowbreak{}latency\_\allowbreak{}mean\_\allowbreak{}ul\_\allowbreak{}ms}, \texttt{e2\_\allowbreak{}du\_\allowbreak{}latency\_\allowbreak{}p90\_\allowbreak{}ul\_\allowbreak{}ms}. & \texttt{e2\_\allowbreak{}du\_\allowbreak{}latency\_\allowbreak{}mean\_\allowbreak{}ul\_\allowbreak{}ms}, \texttt{e2\_\allowbreak{}du\_\allowbreak{}latency\_\allowbreak{}p90\_\allowbreak{}ul\_\allowbreak{}ms}. \\
22 & node & How many KPIs containing ``latency'' are directly connected to the \fnode{}? & 2. & 2. \\
23 & node & How many KPIs containing ``packet\_loss'' are directly connected to the \fnode{}? & 0. & 0. \\
24 & node & How many KPIs containing ``retransmissions'' are directly connected to the \fnode{}? & 0. & 0. \\
\bottomrule
\end{tabular}
\end{table*}

\newpage

\subsection{\system{} Prompt and Triage Report}
\label{subsec:appendix-input-output-agent}

The following Markdown illustration reproduces the full system prompt supplied to \system{} at inference time, defining the O-RAN component hierarchy, the five-step investigative strategy, and the strict report schema the agent must follow (green background illustration). \rev{The subsequent box (yellow background) shows the complete triage report that the agent generated for the incident of Fig.~\ref{fig:causal-graph-e2-delay-low}, reproduced without editing. It lists the two children of $\mathcal{F}$ with their rank and distance (Table~\ref{tab:anomaly-scores-example}), the interface and components confirmed by the answers of Tool~\#2 listed above, a failure chain, a root-cause statement and three recommended diagnostic actions. The agent took $48$\,s for this incident.}

\providecommand{\us}{\_\allowbreak{}}
\begin{myllminputbox}[llminput:perception-agent-prompt]{\system{} prompt.}
You are an expert O-RAN Perception Agent responsible for diagnosing network failures with high precision.

\phead{\#\#\# SYSTEM CONTEXT}
The O-RAN network is a 5G Radio Access Network composed of the following hierarchy:
\begin{promptsteps}
  \item \textbf{Components:}
  \begin{promptitems}
    \item \textbf{O-DU} (Distributed Unit)
    \item \textbf{O-CU} (Centralized Unit)
    \item \textbf{Near-RT RIC} (Near Real-Time RAN Intelligent Controller)
    \item \textbf{Non-RT RIC} (Non-RT RAN Intelligent Controller)
  \end{promptitems}
  \item \textbf{Interfaces:}
  \begin{promptitems}
    \item \textbf{F1-c / F1-u}: Connects O-CU and O-DU (Control/User plane).
    \item \textbf{E2}: Connects Near-RT RIC to O-DU/O-CU.
    \item \textbf{A1}: Connects Non-RT RIC to Near-RT RIC.
  \end{promptitems}
\end{promptsteps}

\phead{\#\#\# YOUR MISSION}
The network has been detected in a failure state. Your task is to diagnose the root cause and map the failure propagation.
Do not guess, hallucinate, or jump to premature conclusions. Adopt a rigorous, \textbf{iterative, investigative strategy}:
\begin{promptsteps}
  \item \textbf{Identify Candidates:} First, use \code{query\us{}root\us{}cause\us{}ranked\us{}list} to find the most likely anomaly source (lowest distance).
  \item \textbf{Verify Topology:} Use \code{query\us{}graph\us{}tool} to confirm which interfaces and downstream components are actually affected by those candidates.
  \item \textbf{Cross-Reference \& Disambiguate:} This is critical. If tool outputs contradict each other, or if anomalies span multiple interfaces (e.g., both F1-c and E2 show issues), you MUST ask further questions using \code{query\us{}graph\us{}tool} to contrast the information.
  \item \textbf{Isolate Root Cause:} Filter out symptomatic noise and confounding variables to identify the true origin point. Do not blindly trust the ranked list, as statistical anomalies can be noisy due to normal network dynamics (e.g., users connecting/disconnecting, load balancing, or routine policy executions). Apply the following strict heuristics:
  \begin{promptitems}
    \item \textbf{Metric Severity Prioritization:} Hard errors carry more diagnostic weight than volume fluctuations. Always prioritize severe degradation metrics (e.g., \code{packet\us{}loss}, \code{retransmissions}, \code{error\us{}rates}) over milder volumetric symptoms (e.g., \code{throughput\us{}kbps}), even if the volumetric metrics have a worse anomaly score or higher ranking.
    \item \textbf{Causal Precedence:} Use the graph tool to determine the direction of the failure. If a severe error on one interface logically precedes a volume drop on another, the severe error is the root cause.
    \item \textbf{Exclude Normal Noise:} Recognize that isolated throughput drops without accompanying hard errors may simply be confounding variables (like user mobility), not the root cause of a network-wide failure.
  \end{promptitems}
  \item \textbf{Synthesize:} Construct the failure chain based ONLY on verified edges and resolved data.
\end{promptsteps}

\phead{\#\#\# AVAILABLE TOOLS}
\begin{promptitems}
  \item \code{query\us{}graph\us{}tool}: Queries the causal graph via natural language. Use this iteratively to check status, verify connections, and contrast conflicting data (e.g., `Is the F1-u interface affected?', `What does node X affect?', `Does anomaly A cause anomaly B?').
  \item \code{query\us{}root\us{}cause\us{}ranked\us{}list}: Returns the top root cause candidates ranked by probability (distance to F-node).
\end{promptitems}
\end{myllminputbox}

\begin{myllminputbox}[llminput:perception-agent-prompt-2]{\system{} prompt.}
\phead{\#\#\# REPORT FORMAT}
Produce a report in the following strict format:

\phead{\# Failure Analysis}

\phead{Root Cause Indicators (Directly Connected to F-node):}
List the top KPIs from the ranked list (include their rank/score).

\phead{Diagnostic Reasoning \& Disambiguation:}
Briefly explain your investigative steps. If multiple interfaces or contradictions were found, explain how you cross-referenced the graph to resolve them.

\phead{Affected Interfaces:}
List interfaces confirmed to be affected (verify status before listing).

\phead{Affected O-RAN Components:}
List components confirmed to be affected based on the interface status.

\phead{Failure Chain:}
A logical explanation of how the failure propagated (e.g., Root KPI $\rightarrow$ Interface $\rightarrow$ Component).

\phead{Root Cause:}
The single most probable technical cause of the failure.

\phead{Recommendation:}
Actionable steps to resolve the specific cause.
\end{myllminputbox}

\begin{myllmoutputbox}[llmoutput:e2-delay-low]{Triage report.}
\phead{\# Failure Analysis}

\phead{Root Cause Indicators (Directly Connected to F-node):}
\begin{promptitems}
  \item \code{e2\us{}du\us{}latency\us{}mean\us{}ul\us{}ms} (Rank 1, Score: 0.0101)
  \item \code{e2\us{}du\us{}latency\us{}p90\us{}ul\us{}ms} (Rank 2, Score: 0.0115)
\end{promptitems}

\phead{Diagnostic Reasoning \& Disambiguation:}
The investigation began by analyzing the root cause ranked list, which strongly indicated that uplink latency on the E2 interface was the primary anomaly. To verify this and rule out symptomatic noise, the causal graph was queried. The graph confirmed that only one interface in the entire network is affected by the anomaly: the \textbf{E2 interface}. Because the F1-c, F1-u, and A1 interfaces are completely unaffected, the failure is isolated to the E2 control path. The direct connection of the F-node to E2 latency KPIs confirms that the latency increase is the origin of the failure, rather than a result of throughput drops or packet loss.

\phead{Affected Interfaces:}
\begin{promptitems}
  \item E2
\end{promptitems}

\phead{Affected O-RAN Components:}
\begin{promptitems}
  \item O-DU
  \item Near-RT RIC
\end{promptitems}

\phead{Failure Chain:}
E2 Uplink Latency (\code{e2\us{}du\us{}latency\us{}mean\us{}ul\us{}ms}/\allowbreak\code{e2\us{}du\us{}latency\us{}p90\us{}ul\us{}ms}) $\rightarrow$ E2 Interface $\rightarrow$ O-DU \& Near-RT RIC.

\phead{Root Cause:}
The root cause is a significant increase in uplink latency on the \textbf{E2 interface}, which is degrading the communication between the O-DU and the Near-RT RIC.

\phead{Recommendation:}
\begin{promptsteps}
  \item Perform a network trace on the E2 interface to identify where the latency is being introduced (e.g., transport network congestion or processing delays).
  \item Check the resource utilization (CPU/Memory) of the O-DU and Near-RT RIC to ensure the latency is not caused by internal processing bottlenecks.
  \item Verify the health of the physical or virtual links connecting the O-DU and the Near-RT RIC.
\end{promptsteps}
\end{myllmoutputbox}

\begin{table*}[ht!]
\centering
\caption{Optuna Search Space and Optimal Hyperparameters for SAM}
\label{tab:sam_hyperparameters}
\begin{tabular}{llcc}
\toprule
\textbf{Parameter} & \textbf{Description} & \textbf{Search Space} & \textbf{Optimal Value} \\
\midrule
\texttt{lambda1} ($\lambda_S$) & Sparsity penalty weight & $[1.0, 50.0]$ (log) & 5.159 \\
\texttt{lambda2} ($\lambda_F$) & Functional complexity penalty & $[0.001, 1.0]$ (log) & 0.046 \\
\texttt{lr} & Generator learning rate & $[10^{-4}, 5 \cdot 10^{-2}]$ (log) & 0.0185 \\
\texttt{dlr} & Discriminator learning rate & $[10^{-4}, 10^{-2}]$ (log) & 0.0037 \\
\texttt{nh} & Generator hidden units & $\{10, 20, 50, 100\}$ & 20 \\
\texttt{dnh} & Discriminator hidden units & $\{100, 200, 350, 500\}$ & 200 \\
\texttt{hlayers} ($H$) & Generator hidden layers & $[1, 3]$ & 2 \\
\texttt{dhlayers} & Discriminator hidden layers & $[1, 3]$ & 1 \\
\texttt{dagstart} & Epoch ratio to begin DAG penalty & $[0.4, 0.7]$ & 0.417 \\
\texttt{dagpenalization\_increase} & Acyclicity penalty increase step & $[0.001, 0.05]$ (log) & 0.041 \\
\midrule
\multicolumn{4}{l}{\rev{\emph{Fixed settings (not searched)}}} \\
\rev{\texttt{train\_epochs}} & \rev{Training epochs} & -- & \rev{3000} \\
\rev{\texttt{test\_epochs}} & \rev{Epochs over which the gates are averaged} & -- & \rev{1000} \\
\rev{$M$} & \rev{SAM runs averaged per incident (seeds $0$--$15$)} & -- & \rev{16} \\
\rev{--} & \rev{Edge threshold on $\bar{A}$ (Sec.~\ref{subsubsec:sam})} & -- & \rev{0.5} \\
\bottomrule
\end{tabular}
\end{table*}

\newpage
\section{Hyperparameters and Training Details}
\label{appendix:hyperparameters}

\subsection{SAM Configurations}
\label{appendix:sam_hyperparameters}

To determine the optimal architecture and training dynamics for the SAM utilized in the causal discovery phase, we conducted an automated hyperparameter search over 110 runs using the Optuna framework\footnote{https://optuna.org/}.

Table \ref{tab:sam_hyperparameters} details the specific hyperparameter search space explored during the optimization trials, alongside the best-performing values adopted for the final model deployment. Consistent with the objective function formulation presented in Sec.~\ref{subsubsec:sam}, \texttt{lambda1} corresponds to the structural sparsity penalty ($\lambda_{S}$), and \texttt{lambda2} corresponds to the functional complexity penalty ($\lambda_{F}$). \rev{The search was carried out for the first version of this work, on earlier recordings of the same testbed, with each graph built from several pooled runs of one attack configuration. The $280$ runs of the revised evaluation (Sec.~\ref{sec:eval}) were recorded afterwards and were not used to select any hyperparameter; we kept the values of Table~\ref{tab:sam_hyperparameters} unchanged. The lower part of the table lists the settings that were fixed rather than searched.} \rev{SAM scales $\lambda_S$ and $\lambda_F$ by the number of samples, so with one incident per graph the two penalties are about five times stronger than during the search; we did not retune them.}

\subsection{Graph Soft-Prompting Encoder Training}
\label{appendix:gin_hyperparameters}

The causal reasoning within the verification phase relies on encoding the graph topological structure into continuous soft tokens that align with the embedding space of a frozen Large Language Model. Based on our configuration, the underlying LLM employed is the \\\texttt{meta-llama/Llama-3.1-8B-Instruct} architecture.

To train the GIN and the linear projection layer, we utilized the Adam optimizer coupled with a linear learning rate warmup schedule. The specific hyperparameter configuration for the graph encoder training is summarized in Table \ref{tab:gin_hyperparameters}.

\begin{table}[H]
\centering
\caption{Hyperparameter Configuration for Graph Soft-Prompting Encoder}
\label{tab:gin_hyperparameters}
\footnotesize
\setlength{\tabcolsep}{4pt}
\begin{tabular}{@{}l>{\raggedright\arraybackslash}p{0.40\columnwidth}c@{}}
\toprule
\textbf{Parameter} & \textbf{Description} & \textbf{Value} \\
\midrule
\texttt{lr} & Peak learning rate (Adam) & $10^{-4}$ \\
\texttt{epochs} & Maximum training epochs & 50 \\
\texttt{warmup\_epochs} & Epochs for \texttt{lr} warmup & 3 \\
\texttt{es\_patience} & Patience for early stopping & 10 \\
\rev{\texttt{batch\_size}} & \rev{Graph--question pairs per step} & \rev{1} \\
\texttt{latent\_dim} & GIN hidden dimension size & 128 \\
\texttt{num\_gte\_layers} & Graph encoder layers & 3 \\
\texttt{gte\_pos\_enc\_size} & LPE size & 4 \\
\texttt{num\_soft\_tokens} & Number of soft tokens & 1 \\
\texttt{readout} & Readout agg. strategy & Graph / Node \\
\rev{\texttt{message\_direction}} & \rev{GIN runs on the reversed graph} & \rev{reverse} \\
\rev{\texttt{use\_edge\_weight}} & \rev{Messages scaled by the gates} & \rev{yes} \\
\rev{\texttt{init}} & \rev{Initial encoder weights (GIN~FT)} & \rev{GraphQA} \\
\rev{\texttt{seed}} & \rev{Training seeds} & \rev{42, 1, 2} \\
\rev{--} & \rev{LLM precision} & \rev{bfloat16} \\
\bottomrule
\end{tabular}
\end{table}

\rev{During data preprocessing, we compute for each graph the positional encodings of Sec.~\ref{subsubsec:encoder}: the eigenvectors of the four smallest strictly positive eigenvalues of the normalized Laplacian of the symmetrized weighted graph, which map directly to the \texttt{gte\_pos\_enc\_size} parameter. During training only, each set of eigenvectors that share an eigenvalue is rotated by a random orthogonal matrix (a random sign flip for a single eigenvector). The generated soft token is prepended to the token embeddings of the text prompt, which contains the node and directed edge list of the graph followed by the natural-language question, while the LLM weights remain strictly frozen. The graph encoder and the projection layer update their parameters through backpropagation by minimizing the loss of Eq.~\eqref{eq:soft-prompt-loss} on the training split only. The validation split is used only for early stopping and to select the checkpoint, and all reported results are computed on the test split (Sec.~\ref{sec:eval}). The encoder is first pre-trained on GraphQA~\cite{fatemi2023talk} and then fine-tuned on O-CIQA, for each readout separately. The agent uses the two checkpoints trained with seed $42$.}

\begin{IEEEbiography}
[{\includegraphics[width=1in,height=1.25in, clip,keepaspectratio]{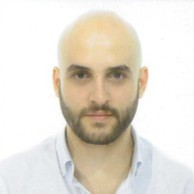}}]{Josep Xavier Salvat Lozano } received his PhD from the Technical University of Kaiserslautern in 2022 and he currently works as a senior research scientist in the 6G Network group at NEC Laboratories Europe, Heidelberg. He worked as a reviewer for several international scientific conferences and journals, including IEEE Transactions on Mobile Computing, IEEE ICC, and Computer Communications Journal and has actively participated in several EU-founded projects, including H2020 5Growth, H2020 Daemon and SNS-JU BeGREEN. His research interests lie in the application of AI to real-world wireless communication systems, including resource allocation and energy-efficiency problems.
\end{IEEEbiography}

\begin{IEEEbiography}[{\includegraphics[width=1in,height=1.25in,clip,keepaspectratio]{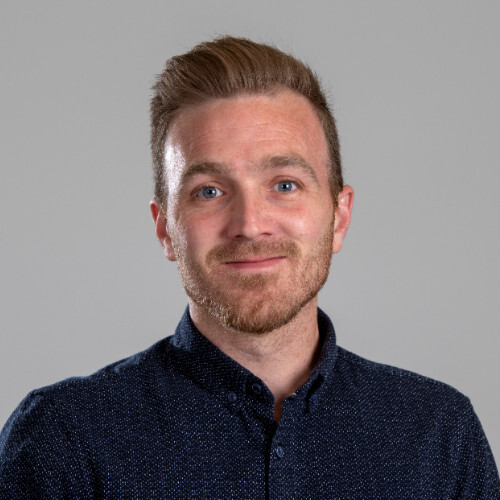}}]{Gines Garcia-Aviles } received a PhD in Telematics Engineering from Universidad Carlos III de Madrid in 2021 and currently works as a senior researcher in the AI-Driven Systems group at the i2CAT Foundation, Barcelona. Gines has served as a reviewer for several international journals and conferences, and as a TPC member of the ACM Open and AI RAN workshop at MobiCom 2026. He has actively participated in several EU-funded projects, including SNS-JU 6G-CLOUD, SNS-JU ORIGAMI and the ESA ARTES project SULFUR. Gines's research interests lie in the reliability, resilience and security of software-based radio access networks, including RAN virtualization and AI-native non-terrestrial 6G networks.
\end{IEEEbiography}

\begin{IEEEbiography}
[{\includegraphics[width=1in,height=1.25in,clip,keepaspectratio]{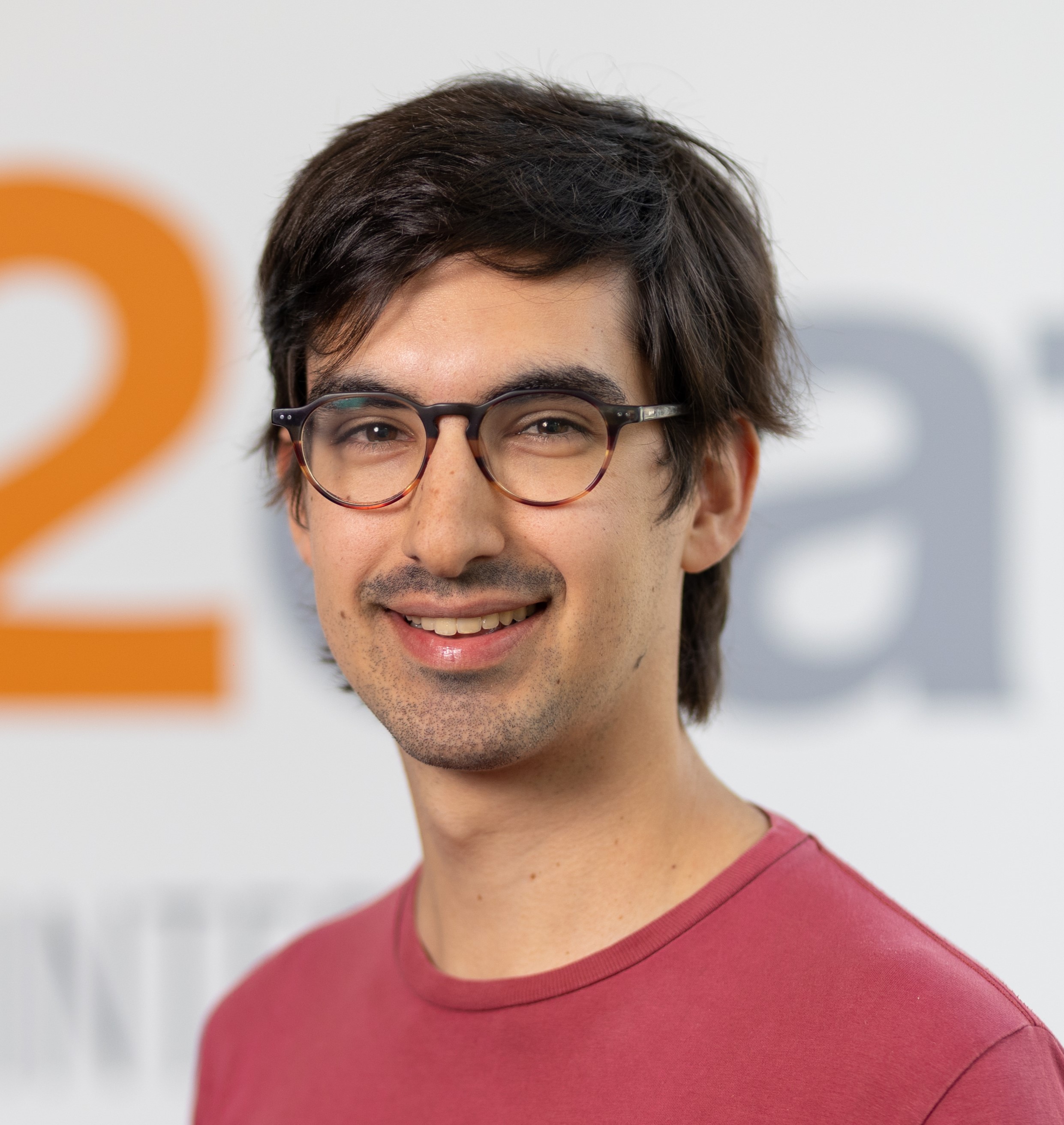}}]{Pau Baguer } 
completed a double engineering degree in aerospace systems and networking from the Polytechnic University of Catalonia (Barcelona) in 2023. He joined the AI-driven Systems group of i2CAT in June 2022 and became a junior researcher in 2023. Pau's research interests are in Open RAN, AI-enhanced systems and UAS.
\end{IEEEbiography}

\begin{IEEEbiography}[{\includegraphics[width=1in,height=1.25in,clip,keepaspectratio]{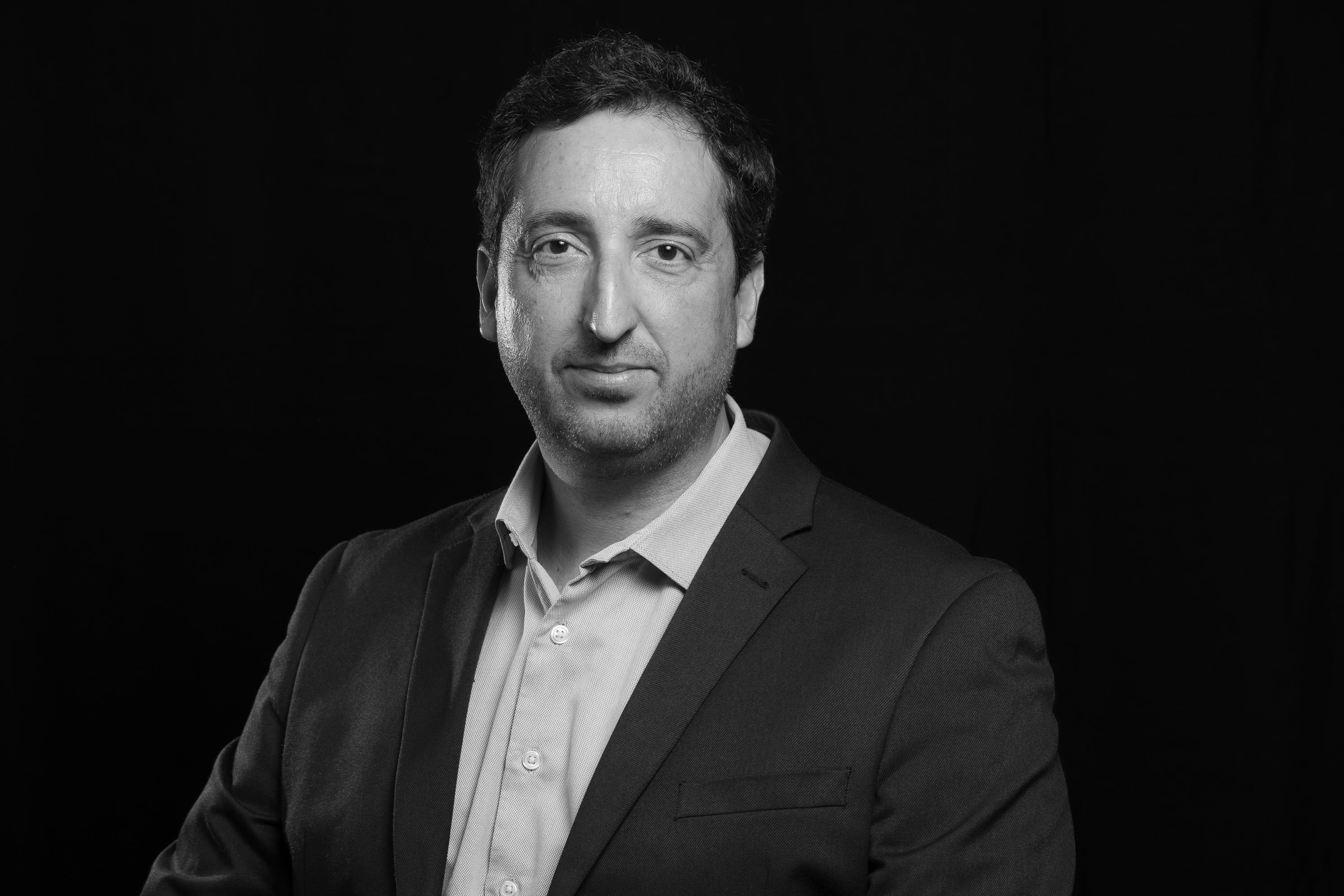}}]{Xavier Costa-Pérez }
is an ICREA Research Professor, Scientific Director at the i2CAT Research Center, and Head of 6G R\&D at NEC Laboratories Europe. He has served on the Organizing Committees of several conferences, published papers of high impact, and holds more than 80 granted patents. He received his PhD degree in telecommunications from the Polytechnic University of Catalonia, Barcelona, and was the recipient of a national award for his PhD thesis.
\end{IEEEbiography}

\end{document}